\newif\ifUsenix\Usenixtrue 
\newif\ifAnon\Anonfalse 

\ifUsenix
  \documentclass[letterpaper,twocolumn,10pt]{article}
  \usepackage{usenix2019_v3}
  \usepackage{authblk}
\else
  \documentclass[conference]{IEEEtran}
\fi
\usepackage{colortbl}
\usepackage{tugraz_defaults}
\usepackage{subfigure}
\usepackage{algorithmicx}
\usepackage[ruled]{algorithm}
\usepackage{algpseudocode} 
\usepackage{multirow}
\usepackage{calc}
\usepackage{tikz-timing}
\usepackage{soul}
\usepackage{xcolor} 

\def\HiLi[#1]{\rlap{\hbox to #1{\color{blue!50}\leaders\hrule height .8\baselineskip depth .5ex\hfill}}}

\algrenewcommand\algorithmicindent{2em}

\newcommand{\Z}{\mathbb{Z}}
\newcommand{\lcm}{\emph{lcm}}

\makeatletter
\algrenewcommand\ALG@beginalgorithmic{\footnotesize}
\makeatother
\algrenewcommand\alglinenumber[1]{\scriptsize #1:}

\algnewcommand{\IIf}[1]{\State\algorithmicif\ #1\ \algorithmicthen}
\algnewcommand{\EndIIf}{\unskip\ \algorithmicend\ \algorithmicif}
\algtext*{EndWhile}
\algtext*{EndIf}
\algtext*{EndFor}
\algtext*{EndProcedure}

\newenvironment{CompactItemize}%
{\begin{list}{$\bullet$}%
  {\leftmargin=\parindent \itemsep=2pt \topsep=2pt
    \parsep=0pt \partopsep=0pt}}%
{\end{list}}

\newcommand{\para}[1]{\vspace{-.5em}\paragraph{{#1.}}}

\makeatletter
\newcommand\footnoteref[1]{\protected@xdef\@thefnmark{\ref{#1}}\@footnotemark}
\makeatother

\PassOptionsToPackage{hyphens}{url}\usepackage{hyperref}
    \definecolor{linkcolor}{rgb}{0,0.1,0.35}
    \definecolor{citecolor}{rgb}{0,0.4,0}
    \definecolor{urlcolor}{rgb}{0,0,0.65}
    \hypersetup{pdfpagemode=UseNone,pdfstartview=FitH,colorlinks=true, linkcolor=linkcolor, urlcolor=urlcolor, citecolor=citecolor}

\usepackage{enumitem}
\newlist{paraenum}{enumerate*}{1}
\setlist[paraenum]{label=\emph{(\roman*)}}

\usepackage{mdframed}
\newmdenv[
  leftmargin = 0pt,
  innerleftmargin = 1em,
  innertopmargin = 0pt,
  innerbottommargin = 0pt,
  innerrightmargin = 0pt,
  rightmargin = 0pt,
  linewidth = 3.5pt,
  linecolor= gray!50,
  topline = false,
  rightline = false,
  bottomline = false
  ]{myleftbar}

\begin{document}

\newcommand{\attackName}{CopyCat} 
\newcommand{\attack}{{\textsc{\attackName}}\xspace} 
\newcommand{\primeprobe}{{\textsc{Prime+Probe}}\xspace} 
\newcommand{\flushreload}{{\textsc{Flush+Reload}}\xspace} 
\newcommand{\flushflush}{{\textsc{Flush+Flush}}\xspace} 

\title{
   \Large \bf \attack:
   Controlled Instruction-Level Attacks on Enclaves
}

\ifAnon  
  \author{Anonymous Submission}
  \ifUsenix     
  \else    
    \IEEEspecialpapernotice{}
  \fi
\else
  \ifUsenix
    \author[1]{Daniel Moghimi}
    \author[2]{Jo Van Bulck}
    \author[3]{Nadia Heninger}
    \author[2]{Frank Piessens}
    \author[1]{Berk Sunar}
    \affil[1]{Worcester Polytechnic Institute, Worcester, MA, USA}
    \affil[2]{imec-DistriNet, KU Leuven, Leuven, Belgium}
    \affil[3]{University of California, San Diego, CA, USA}
    \else
    \author{
      \IEEEauthorblockN{
        Daniel Moghimi\IEEEauthorrefmark{1},
    }

    \IEEEauthorblockA{\IEEEauthorrefmark{1}Worcester Polytechnic Institute}
  \fi
\fi

\maketitle

{
\noindent\textbf{This
paper will be presented at USENIX Security Symposium 2020.
Please cite this work as:}
\begin{myleftbar}
\emph{Daniel Moghimi, Jo Van Bulck, Nadia Heninger, Frank Piessens, Berk Sunar,
\enquote{CopyCat: Controlled Instruction-Level Attacks on Enclaves} in Proceedings of the 
29th USENIX Security Symposium, Boston, MA, August 2020.}
\end{myleftbar}
}

\begin{abstract}  
The adversarial model presented by trusted execution environments (TEEs) has 
prompted researchers to investigate unusual attack vectors. 
One particularly powerful class of \emph{controlled-channel} attacks abuses 
page-table modifications to reliably track enclave memory accesses at a page-level granularity. 
In contrast to noisy microarchitectural timing leakage, this line of 
deterministic controlled-channel attacks abuses indispensable architectural 
interfaces and hence cannot be mitigated by tweaking microarchitectural resources.

We propose an innovative controlled-channel attack, named \attack, that 
deterministically {counts} the number of instructions executed \emph{within} a single enclave code page. 
We show that combining the instruction counts harvested by \attack with traditional, 
coarse-grained page-level leakage allows the accurate reconstruction of enclave control 
flow at a \emph{maximal} instruction-level granularity.
\attack can identify intra-page and intra-cache line branch decisions that ultimately may 
only differ in a single instruction, underscoring that even extremely 
subtle control flow deviations can be deterministically leaked from secure enclaves.
We demonstrate the improved resolution and practicality of \attack on Intel SGX 
in an extensive study of single-trace and deterministic attacks against 
cryptographic implementations, and give novel algorithmic attacks 
to perform single-trace key extraction that exploit subtle vulnerabilities in the 
latest versions of widely-used cryptographic libraries. Our findings highlight the 
importance of stricter verification of cryptographic implementations, especially 
in the context of TEEs.
\end{abstract}

\section{Introduction}\label{sec:intro}

In the past years, we have seen a continuous stream of software-based side-channel attacks~\cite{andrysco2015subnormal,
evtyushkin2018branchscope,yarom2017cachebleed,moghimi2018memjam,aldaya2018port,yarom2014flush,ge2016survey}.
A first category of {microarchitectural timing} attacks commonly abuses optimizations in modern processors,
where secret-dependent state is accumulated in various microarchitectural buffers during the victim's execution.
If these buffers are not flushed before a context switch to an attacker domain, 
victim secrets can be reconstructed by observing timing variations by the attacker.
The success of these attacks critically relies on subtle timing differences,
making them inherently non-deterministic and prone to measurement noise~\cite{ge2016survey}.
Usually, this class of stateful attacks can be eliminated by isolating leaky 
microarchitectural resources~\cite{liu2016catalyst,townley2019smt,xu2017vcat,dessouky2019hybcache}.

Orthogonal to the first class of microarchitectural timing attacks, 
recent research on \emph{controlled-channel} attacks~\cite{xu2015controlled,van2017telling,wang2017leaky,gyselinck2018off} has 
abused the processor's privileged software interface to extract fully deterministic, 
noise-free side-channel access patterns from enclave applications.
While the operating system (OS) was traditionally not considered to be under the attacker's control,
this assumption fundamentally changed with the rise of trusted execution environments (TEEs), such as Intel SGX.
Prior research~\cite{xu2015controlled,van2017telling} has identified page-table accesses 
and faults as privileged interfaces that can be exploited as no-noise 
controlled channels to deterministically reveal enclave memory accesses at a 4\,KiB page-level granularity.
The paging channel has drawn considerable research attention since it abuses an 
intrinsic property of the x86 processor architecture without relying on 
microarchitectural states. 
In particular, controlled-channel attacks have proven to be challenging to 
mitigate in a principled way, in spite of numerous defense proposals~\cite{shinde2016preventing,chen2017detecting,shih2017t,chenracing,strackx2017heisenberg,sasy2017zerotrace,orenbach2020autarky}.

In this work, we show that the resolution of deterministic controlled-channel attacks 
extends well beyond the relatively coarse-grained 4\,KiB page-level granularity.
We introduce \attack, an innovative \emph{interrupt-counting} channel that can precisely 
reconstruct the intra-page control flow of a secure enclave at a maximal, instruction-level granularity.
Our attack leverages the SGX-Step~\cite{vanbulck2017SGXStep} framework to forcibly 
step into a victim enclave code exactly one instruction at a time.
While high-frequency timer interrupts have previously been leveraged to boost 
microarchitectural timing attacks~\cite{van2018nemesis,moghimi2017cachezoom,hahnel2017high,lee2017inferring,huo2020bluethunder}, 
we exploit the architectural interrupt interface itself as a deterministic controlled channel.
In short, our attacks rely on the key observation that merely counting the number
of times a victim enclave can be interrupted directly reveals the number of
executed instructions.
We show that combining our fine-grained interrupt-based counting technique with traditional, 
coarse-grained page-table access patterns~\cite{van2017telling,wang2017leaky} as a secondary oracle allows us to construct highly effective and deterministic attacks that track enclave control flow 
at a maximal, {instruction-level} granularity.
Crucially, the improved temporal dimension of \attack overcomes the spatial resolution limitation of prior controlled-channel attacks, invalidating a key assumption in 
some prior defenses~\cite{shinde2016preventing,intelsgxdev} that presumes that adversaries 
can only deterministically monitor enclave memory accesses at a coarse-grained 4\,KiB granularity.
Furthermore, in contrast to previous high-resolution SGX side channels~\cite{lee2017inferring,van2018nemesis,aldaya2018port,moghimi2018memjam,moghimi2017cachezoom} 
that rely on timing differences from contention in some shared microarchitectural state,
\attack cannot be transparently mitigated by isolating microarchitectural resources.

To demonstrate the strength of \attack, we develop single-trace attacks that allow efficient cryptographic key recovery from multiple widely-used cryptographic libraries.  We extend the cryptanalysis of the binary Euclidean algorithm, which is used for modular inversion in most of the common libraries we examined, and give novel algorithms for efficiently recovering cryptographic keys from a single control flow trace for DSA and ECDSA digital signature generation and RSA key generation.  The libraries we examined implemented numerous mitigations against side-channel attacks, including always-add-and-double for elliptic curve scalar multiplication and RSA exponent masking, but these protections were insufficient to protect against \attack.  We conclude that new classes of defenses will be necessary to protect against this type of high-granularity, deterministic, and noise-free attack.

\vspace{-.7em}
\para{Contributions}
In summary, our main contributions are:

\begin{CompactItemize}
  \item We propose \attack: a novel deterministic controlled-channel attack to leak runtime 
  control flow from Intel SGX enclaves without noise at an instruction-level granularity.
  
  \item We explore the impact of \attack on non-crypto applications by defeating
  a state-of-the-art compiler hardening technique against branch shadowing attacks.
  
  \item In an extensive empirical case study of side-channel vulnerabilities 
  in widely-used cryptographic libraries including WolfSSL, Libgcrypt, OpenSSL, 
  and Intel IPP, we verify the practicality and capability of these attacks, demonstrate 
  several attacks, and report vulnerabilities in some of these libraries.

  \item We devise new algorithmic techniques to exploit these vulnerabilities 
  in DSA, ECDSA, and ElGamal, as well as RSA key generation, which result in 
  complete key recovery in the context of Intel SGX.

  \item Finally, we outline requirements and pitfalls for countermeasures and 
  mitigations in hardware and software.

\end{CompactItemize}

\vspace{-.7em}
\para{Responsible Disclosure}
We reported the weaknesses in WolfSSL in Nov. 2019 and provided guidelines 
for mitigation, tracked via CVEs 2019-1996{\{0,1,3\}} and CVE-2020-7960.
We reported our findings to OpenSSL and Libgcrypt teams in Feb. 2020.
OpenSSL replaced \texttt{BN\_gcd} with a constant-time 
implementation~\cite{bernstein2019fast} in version 1.1.1e. 
Libgcrypt issued a similar fix that will appear in version 1.8.6.

We shared our attack with the Intel product security incident response team (iPSIRT), 
who acknowledged that \attack leaks side-channel information, but re-iterated that protecting 
against side channels requires the enclave developer to follow the constant-time 
coding best practices as advised by Intel~\cite{IntelCst}.
\Cref{sec:countermeasure} elaborates further on mitigations and explains how
fully preventing our attacks requires the meticulous application of 
constant-time programming paradigms.

\section{Background and Related Work}\label{sec:back}
\subsection{Side-Channel Attacks on Intel SGX}

Recent Intel processors include software guard extensions (SGX)~\cite{inteSDM} to allow trusted execution of critical code in so-called \emph{enclaves} on top of a potentially compromised OS.
SGX enclaves are isolated at runtime in a memory area that is transparently encrypted and can be remotely attested by the processor.
Dedicated \texttt{eenter} and \texttt{eexit} instructions switch the processor in and out of \enquote{enclave mode}.

Importantly, while the confidentiality and integrity of enclaved execution is always safeguarded by the processor, 
traditionally privileged OS software remains in charge of availability concerns.
SGX enclaves live in the virtual address space of a conventional, user-space process.
To allow for demand-paging and oversubscription of the physically available
encrypted memory, enclave page-table mappings are verified but remain under the
explicit control of the untrusted OS.
Recent address translations may be cached in an internal translation
lookaside buffer (TLB), which is flushed by the processor on every enclave transition.
When delivering asynchronous interrupts or exceptions, the processor takes care
to securely save and scrub CPU registers before exiting the enclave, which can
be subsequently re-entered through the \texttt{eresume} instruction.
Furthermore, in case of a page-fault event, the processor clears the lower bits
representing the page offset in the reported address to ensure that the OS can
only observe enclave memory accesses at a 4\,KiB page-level granularity.

\begin{table*}[ht]
  \small
  \centering
  \caption{Characterization of demonstrated Intel SGX microarchitectural side channels (top) and controlled channels (bottom).
          Our novel \attack technique is highlighted at the bottom and combines noise-free interrupt counting measurements with deterministic page table accesses to reconstruct enclave-private control flow at a maximal, instruction-level granularity.
          }
  \medskip
  \label{tab:sgx-attack}
  \adjustbox{max width=.95\linewidth}{
  \begin{tabular}{lllll}
  \toprule
  & \textbf{Attack} & \textbf{Data} & \textbf{Granularity} & \textbf{Noise} \\
  \midrule
  \multirow{6}{*}{\rotatebox{90}{\textsl{$\mu$-arch contention}}}
  & DRAM row buffer conflicts~\cite{wang2017leaky}            & Code + data & \xmark Low (1-8\,KiB) & \xmark High \\
  & \primeprobe cache conflicts~\cite{moghimi2017cachezoom,hahnel2017high,brasser2017software,schwarz2017malware}         & Code + data & \xmark Med (64-512\,B cache line/set) & \tmark Med \\
  & Read-after-write false dependencies~\cite{moghimi2018memjam} & Data & \cmark High (4\,B) & \xmark High \\
  & Branch prediction history buffers~\cite{lee2017inferring,evtyushkin2018branchscope,huo2020bluethunder} & Code & \cmark High (branch instruction) & \tmark Low \\
  & Interrupt latency~\cite{van2018nemesis}                   & Code + data & \cmark High (instruction latency class) & \xmark High \\
  & Port contention~\cite{aldaya2018port}                     & Code        & \cmark High ($\mu$-op execution port)    & \xmark High \\
  \midrule
  \multirow{4}{*}{\rotatebox{90}{\textsl{Ctrl channel}}}
  & Page faults~\cite{xu2015controlled} and page table A/D bits~\cite{van2017telling,wang2017leaky}                      & Code + data & \xmark Low (4\,KiB ) & \cmark Deterministic \\
  & IA-32 segmentation faults~\cite{gyselinck2018off}         & Code + data & \xmark Low/high (4\,KiB; 1\,B for enclaves $\le$ 1\,MiB) & \cmark Deterministic \\
  & Page table \flushreload~\cite{van2017telling}             & Code + data & \xmark Low (32\,KiB) & \tmark Low \\
  & \cellcolor{orange!25}\bf \attack & \cellcolor{orange!25}\bf Code  & \cellcolor{orange!25}\cmark \bf High (instruction) & \cellcolor{orange!25}\cmark \bf Deterministic \\
  \bottomrule
  \end{tabular}
  }
\end{table*}

While Intel SGX provides strong architectural isolation, several studies have highlighted that enclave secrets may still leak through side-channel analysis.
\Cref{tab:sgx-attack} summarizes how all previously demonstrated side-channel attacks fall into two categories:\footnote{
    Transient-execution attacks~\cite{van2018foreshadow,schwarz2019zombieload,vanbulck2020lvi} are orthogonal to 
    metadata leakage through side channels and require recovery of the trusted computing base through complementary microcode and compiler mitigations.
}
\begin{paraenum}
    \item microarchitectural timing attacks, which may achieve a high granularity 
but are inherently prone to measurement noise, and
    \item fully deterministic controlled-channel attacks that 
only offer a relatively coarse grained 4\,KiB page-level granularity.
\end{paraenum}
\attack proposes the only generally applicable controlled-channel attack 
that is both fully deterministic and offers a maximal, instruction-level granularity.

\para{Microarchitectural Contention}
Microarchitectural timing side-channel attacks exploit the fact that various resources, such as caches~\cite{moghimi2017cachezoom,hahnel2017high,brasser2017software,schwarz2017malware}, 
DRAM row buffers~\cite{wang2017leaky}, branch predictors~\cite{lee2017inferring,evtyushkin2018branchscope,huo2020bluethunder}, 
dependency resolution logic~\cite{moghimi2018memjam}, or execution ports~\cite{aldaya2018port} are competitively shared 
between sibling CPU threads or not flushed when exiting the enclave.
This contention causes measurable timing differences in the attacker domain, 
allowing the attacker to infer the private control flow or data access pattern of the enclave with varying degrees of granularity.
In the context of a TEE such as Intel SGX, such attacks can be mounted with less noise 
and improved resolution because the adversary controls the OS.

In particular, one line of work has developed interrupt-driven 
attacks~\cite{moghimi2017cachezoom,hahnel2017high,lee2017inferring,vanbulck2017SGXStep} 
that rely on frequent enclave preemption to sample side-channel measurements at an 
improved temporal resolution. This technique has been demonstrated to amplify
side-channel leakage from the cache~\cite{moghimi2017cachezoom}, 
the branch target buffer~\cite{lee2017inferring}, and the directional branch predictor~\cite{huo2020bluethunder}.
Similar techniques have been applied to attack ARM TrustZone~\cite{ryan2019hardware}.
Nemesis~\cite{van2018nemesis} showed that while single stepping, the response time 
to service an interrupt may reveal which instruction is being executed in the pipeline.
The SGX-Step framework~\cite{vanbulck2017SGXStep} has been leveraged in several 
other microarchitectural attacks~\cite{van2018nemesis,huo2020bluethunder,cabrera2020when,van2018foreshadow,schwarz2019zombieload,vanbulck2020lvi}
to reliably single-step enclaves at a \emph{maximal} temporal resolution by means of precise and short timer interrupt intervals.

\para{Controlled-Channel Attacks}
Xu \etal\cite{xu2015controlled} first showed how privileged adversaries can revoke access 
rights on a specific enclave page and be deterministically notified 
by means of a page-fault signal when the enclave next accesses that page.
They demonstrated several attacks on non-cryptographic applications by observing
that page-fault sequences uniquely identify specific points in the victim's execution.
Subsequent work~\cite{van2017telling,wang2017leaky} developed stealthier techniques 
to extract the same information without provoking page faults.
These attacks interrupt the victim enclave to forcefully flush the TLB and
provoke page-table walks, which can later be reconstructed through \enquote{accessed}
and \enquote{dirty} attributes or cache timing differences for untrusted page-table entries.
Finally, Gyselinck \etal\cite{gyselinck2018off} demonstrated an alternative 
controlled-channel attack that abuses legacy IA32 segmentation faults.
Their attack offers an improved, byte-level granularity in the first MiB of the 
enclave address space, but only for the unusual case of a 32-bit enclave, and 
this behavior has since been fixed in recent microcode.

With \attack, we contribute an improved attack technique to refine the resolution of 
existing controlled channels by precisely counting the number of executed
enclave instructions between successive page accesses.
Prior work has similarly suggested an additional temporal dimension for the paging
channel by using interrupts to reconstruct \texttt{strlen} loop
iterations~\cite{vanbulck2017SGXStep,van2019tale}, or by logging noisy
wall-clock time~\cite{wang2017leaky} for page-access
events to improve stealthiness and reduce the number of TLB flushes.
Recent work~\cite{kim2019sgx} on enclave control flow obfuscation furthermore 
investigated using single-stepping in an SGX simulator to probabilistically 
identify software versions in an emulated enclave debug environment.
In contrast to these specialized cases, \attack explicitly
recognizes instruction counting as a practical and generically applicable
attack primitive that can deterministically capture the execution trace within a
single enclave code page.

\subsection{Cryptographic Signature Schemes}\label{sec:publickey}
Signature schemes are extensively used for remote attestation and 
authentication of trusted enclaves such as Intel SGX~\cite{intelsgxdev}. Moreover, 
TEEs like Intel SGX can promise trusted execution of these algorithms for a wide 
range of applications such as trusted key management~\cite{fortanix} and private contact 
discovery~\cite{signal}. In this section, we provide an overview of signing algorithms
based on public-key cryptography (PKC) that are used in our attack demonstrations. 

\para{RSA}
RSA keys~\cite{rivest1978method} are generated 
as follows:
\begin{compactenum}
 \item Choose large prime numbers $p$ and $q$, compute $N = pq$,
 \item Compute the least common multiple $\lambda(N) = \lcm(p\text{-}1, q\text{-}1)$,
 \item Choose $e$ such that $1 < e < \lambda(N)$ and $gcd(e, \lambda(N)) = 1$,
 \item Compute $d = e^{-1} \bmod \lambda(N)$.
\end{compactenum}
$(N, e)$ are public and $(p, q, \lambda(N), d)$ are private.
RSA implementations commonly use the Chinese remainder theorem (CRT) to reduce computation time, and generate additional private values $d_P= d \bmod (p-1)$, $d_Q= d \bmod (q-1)$, 
and $q_{inv} = q^{-1} \bmod p$.
A signature is the value $s = h^d \bmod\ N$ where $h$ is a hashed and padded message. 
Signature verification checks if $h \equiv s^e \bmod N$.
To prevent side-channel attacks on signature generation, most implementations
blind the input $h$ with a random $r$ before computing the modular 
exponentiation: $s_b = (h r^e)^d \bmod N = h^d r \bmod N$. Later, the unblinded 
signature can be computed as $s = s_b r^{-1} \bmod N$. 
As a result, attacks on RSA key generation have gained recent
attention~\cite{aldaya2017spa,cabrera2020when}. 
However, since the private key parameters are only computed once, an attack against
RSA key generation must only require a single trace.

\para{DSA and ElGamal}
In the Digital Signature Algorithm (DSA)~\cite{gallagher2013digital},
the public parameters are a prime $p$, another prime divisor $n$ of $p - 1$, and the group generator $g$. 
The private key $x$ is chosen randomly such that $1 < x < n-1$, and the public key is $y = g^x (\bmod p)$.
To sign a message hash $h$:
\begin{compactenum}
 \item Choose a random secret $k$ such that $1 < k < n-1$,
 \item Compute $r = g^k \bmod p \bmod n$,
 \item Compute $s = k^{-1}(h + r \cdot x) \bmod n$.
\end{compactenum}
$(r, s)$ is the output signature pair.

In the ElGamal signature scheme, an alternative to DSA, 
the first signature pair $r$ is computed similarly, but 
the second pair is computed as $s=k^{-1}(h-r \cdot x) \bmod (p-1)$.

\para{ECDSA}
Elliptic-curve DSA (ECDSA) is similar to DSA.
The public parameters are an elliptic curve $E$ with scalar multiplication operation $\times$, a point $G$ on the curve, and the integer order $n$ of $G$ over $E$. 
The secret key $d$ is a random integer satisfying $1 < d < n-1$, and the public key is $Q = d\times G$.
Signature generation for a message hash $h$ is as follows:
\begin{compactenum}
 \item Choose a random secret $k$ such that $1 < k < n-1$,
 \item Compute $(x, y) = k \times G$ and $r = x \bmod n$,
 \item Compute $s = k^{-1}(h + r \cdot d) \bmod n$.
\end{compactenum}
$(r, s)$ is the output signature pair.

In DSA, ECDSA and ElGamal, it is critical for $k$ to be uniquely chosen for each
signature generation and to remain secret. 
Exposing one instance of $k$ for a known signature results in a simple key recovery: $d = r^{-1}(s \cdot k - h) \bmod n$.
Since $k$ is an ephemeral value, a noisy side-channel attack against $k$ cannot 
reduce the sampling noise using multiple runs of the attack.  
However, as discussed in \cref{sec:relatedpkcscsa}, lattice attacks can recover 
the signing key from partial knowledge of $k$ for many signatures.
In~\cref{sec:wolfdsa} and \cref{sec:crypto}, we show that we can recover 
the entire ephemeral $k$ deterministically in a single trace of the 
computation of the modular inverse $k^{-1} \bmod n$.
Single-trace attacks on signature generation illustrate vulnerabilities even in scenarios where an attacker cannot trigger multiple signature generation operations or can only collect a single trace.

\subsection{Side-Channel Attacks on PKC Schemes}
\label{sec:relatedpkcscsa}
Public-key algorithms that execute variable operations
for each bit of a secret input, like the square-and-multiply algorithm for modular exponentiation, 
and scalar multiplication based on Montgomery ladders, are susceptible to side-channel leakage. Such algorithms 
have been exploited in naive attacks~\cite{zhang2012cross,yarom2014flush,yarom2014recovering} 
where the victim is triggered many times to compensate for potential sampling noise. 
These attacks generally conclude with the recovery of most of the secret bits.
Nowadays, most implementations have adopted constant-time algorithms like
fixed-window scalar multiplication to mitigate such attacks~\cite{rivain2011fast}.

\para{Key Recovery using Partial Information}
Key recovery from DSA and ECDSA with partial knowledge of the nonce $k$ can be 
solved efficiently using lattices~\cite{boneh1996hardness, nguyen2003insecurity}.
These attacks apply to the case when a few bits are leaked about the nonce for multiple signatures, and the adversary can sample many signatures.
Researchers have applied lattice-based attacks to non-constant time algorithms
that leak some information about $k$~\cite{pereida2016make,benger2014ooh,ryan2019return}.
Garcia \etal~\cite{garcia2017constant} demonstrate 
an attack that recovers the sequence of divisions and subtractions from the 
binary extended Euclidean algorithm (BEEA) for modular inversion. 
They observe that this sequence leaks some least significant bits of $k$ 
and apply a lattice-based key recovery algorithm. In contrast, \attack 
allows full key recovery from a single DSA signature trace,
 even for a compact BEEA implementation (\S\ref{sec:wolfdsa}). 
We generalize this attack to another vulnerable modular inverse implementation 
used for DSA, ECDSA, and ElGamal (\S\ref{sec:crypto}).

Even subtle implementation flaws that leak the bit length of $k$ are sufficient 
for multi-trace lattice-based key recovery~\cite{brumley2011remote,moghimi2019tpm,dall2018cachequote}.
In these cases, while the algorithm was implemented with enough care to avoid 
secret-dependent conditional statements, they leak the bit length by skipping 
the most significant zero bits of $k$. 
In \cref{sec:wolfecdsa}, we exploit a countermeasure against this attack to precisely
leak the nonce length, and recover the secret key using a lattice attack.

\para{Single-Trace Attacks on RSA}
Recent work has demonstrated a single-trace side-channel attack against RSA key generation 
that leaks the sequence of divisions and subtractions
from the BEEA during the coprimality test $\gcd(e, p-1)$~\cite{weiser2018single,aldaya2019cache} or secret key generation $d = e^{-1} \bmod \lambda(N)$~\cite{cabrera2017side}. 
These attacks recover the secrets $(p-1)$ or $\lcm(p-1, q-1)$ from this sequence when $e$ is small enough to be brute forced, which is typically the case in practice\footnote{$e$ is commonly chosen as $2^{16} + 1 = 65537$.}.
The proposed mitigation is to increase the size of the input $e$ 
by masking it with a random variable that may be hard coded~\cite{cabrera2017side}. 
In \cref{sec:wolfrsa}, we use \attack to recover all the branches from BEEA, 
not just the sequence of divisions and subtractions.
We propose a novel algorithm that uses this information to recover the private 
factors $p$ and $q$ from $e^{-1} \bmod \lambda(N)$. Our attack works even for 
large $e$, thwarting the above mitigations. 

Furthermore, our algorithm is even able to recover the key from a modular 
inversion algorithm with multiple unknowns. 
We demonstrate a novel end-to-end single-trace attack on the CRT computation $q^{-1} \bmod p$. 
In a concurrent and independent work, Aldaya \etal\cite{cabrera2020when} outline a different
key recovery algorithm for $q^{-1} \bmod p$ that is not always successful.
Our single-trace attacks on RSA in \cref{sec:wolfrsa} use a
branch-and-prune algorithm inspired by Heninger and Shacham~\cite{heninger2009reconstructing}.
Bernstein \etal applied a variant of branch-and-prune algorithm to recover 
RSA keys from a sliding-window modular exponentiation implementation~\cite{bernstein2017sliding}.
Similarly, Yarom \etal demonstrated an attack with intra-cache 
line granularity on a fixed-window implementation of modular exponentiation 
that recovers a fraction of the bits~\cite{yarom2017cachebleed}.
In \cref{sec:crypto}, we generalize our attack to implementations of BEEA used in other popular 
cryptographic libraries. We demonstrate attacks against $gcd(p-1, q-1)$ in OpenSSL X.931 RSA and 
$q^{-1} \bmod p$ and $e^{-1} \bmod \lambda(N)$ in WolfSSL and Libgcrypt. 

\section{\attack Attack} \label{sec:method}

\para{Attacker Model}
We assume the standard Intel SGX root adversary model with full control over the untrusted OS~\cite{inteSDM}.
SGX's strong threat model is justified, for instance, by considering 
untrusted cloud providers under the jurisdiction of foreign nation states, or
end users with an incentive to break DRM technology running on their own device.
Following prior work, we assume a remote, software-only adversary
who has compromised the untrusted OS, allowing the x86 APIC timer
device to be configured to precisely interrupt the
enclave~\cite{vanbulck2017SGXStep,moghimi2017cachezoom,hahnel2017high,lee2017inferring}
and modify page-table entries to learn enclaved memory accessed at a 4\,KiB
granularity~\cite{xu2015controlled,shinde2016preventing,van2017telling}.
Like previous attacks, we further assume knowledge of the victim application,
either through source code or the application binary.
We assume the enclave code is free from memory-safety vulnerabilities~\cite{van2019tale} and the 
Intel SGX platform is properly updated against transient-execution 
attacks~\cite{van2018foreshadow,schwarz2019zombieload}.

The adversary's goal is to learn fine-grained control-flow decisions in the victim enclave.
In contrast to noisy microarchitectural side channels~\cite{brasser2017software,moghimi2017cachezoom,moghimi2018memjam,lee2017inferring,van2018nemesis,aldaya2018port},
we can also target victims who process a secret only \emph{once} in a {single run} 
(as is the case in key generation) and hence victims who cannot be forced to repeatedly
perform computations on the same secret multiple times.
Crucially, in contrast to prior controlled-channel attacks~\cite{xu2015controlled,van2017telling}, 
\attack offers intra-page granularity and we assume that conditional control flow blocks in the victim 
enclave are aligned \enquote{to exist entirely within a single page} as officially recommended by Intel~\cite{intelsgxdev}.

\subsection{Building the Interrupt Primitive}
\label{sec:fusion}

Debug features like the x86 single-step trap flag are explicitly disabled 
by the Intel SGX design~\cite{inteSDM} while in enclave mode.
Recent research, however, has demonstrated that root adversaries 
may abuse APIC timer interrupts to forcibly pause a victim enclave at fixed time intervals.
We build our interrupt primitive on top of the open-source SGX-Step~\cite{vanbulck2017SGXStep} 
framework, which offers a maximal temporal resolution by reliably interrupting the victim 
enclave at most one instruction at a time.
SGX-Step comes in the form of a Linux kernel driver and runtime library to configure APIC 
timer interrupts and untrusted page-table entries directly from user space.

\para{Deterministic Single-Stepping}
We first establish a suitable value for the platform-specific \texttt{SGX\_STEP\_TIMER\_INTERVAL} 
parameter using the SGX-Step benchmark tool on our target processor.
This ensures that the victim enclave always executes {at most} one instruction at a time.
Previous studies~\cite{vanbulck2017SGXStep,van2018nemesis,huo2020bluethunder} have 
reported reliable single-stepping results with SGX-Step for enclaves with several hundred thousand instructions where 
in the vast majority of cases ($>97\%$) the timer interrupt arrives within the \emph{first} 
enclave instruction after \texttt{eresume}, \ie single-step, and in \emph{all} other cases 
the interrupt arrives within \texttt{eresume} itself, \ie zero-step before an enclave instruction is ever executed.
Furthermore, zero-step events can be filtered out by observing that the \enquote{accessed} bit in 
the untrusted page-table entry mapping the enclave code page is only ever set by the processor 
when the interrupt arrived after \texttt{eresume} and the enclave instruction has indeed been retired~\cite{van2018nemesis}.
Hence, to achieve {noiseless and deterministic} single-stepping for revealing code and 
data accesses at an instruction-level granularity,
we rely on the observation that a properly configured timer \emph{never} causes a multi-step, 
and we discard any zero-step events by querying the \enquote{accessed} bit in
the untrusted page-table entry mapping the current enclave code page.
The experimental evaluation in \cref{sec:wolfssl} confirms that our single-stepping interrupt primitive 
indeed behaves fully deterministically when using \attack to count several
millions of enclave instructions.

Before entering single-stepping mode, we first use a coarse-grained page-fault
state machine to easily advance the enclaved execution to a specific function
invocation on the targeted code page.
Such page-fault sequences have priorly been shown to uniquely locate specific
execution points in large binaries~\cite{xu2015controlled,shinde2016preventing,weiser2018single}.
Once the specific code page of interest has been located, \attack starts
counting instructions until detecting the next code or data page access 
to reveal instruction-level control flow.

\para{Effects of Macro Fusion}
Interestingly, we found that \attack can also be used to study a microarchitectural optimization in 
recent Intel Core processors, referred to as \emph{macro fusion}~\cite{inteloptimze,wikichipmacro}.
The idea behind this optimization technique is to combine certain adjacent instruction pairs in 
the front-end into a single micro-op that executes with a single dispatch and hence frees up space in the processor pipeline.

Intel documents that fusion only takes place for some well-defined
compare-and-branch instruction pairs~\cite[\S3.4.2.2]{inteloptimze},
which are additionally not split on a cache line
boundary~\cite[\S2.4.2.1]{inteloptimze}.
We experimentally found that for fusible instruction pairs, 
\attack consistently counts only \emph{one} interrupt, even though the enclave-private program counter 
has been advanced with \emph{two} assembly instructions forming the fused pair.
Our experimental observations on Kaby Lake confirm Intel's documented limitations,
\eg \texttt{test;jo} can be fused (interrupted once) but not \texttt{cmp;jo} (interrupted twice); and 
fusible pairs that are split across an exact cache line boundary are not fused (interrupted twice).
Importantly, we found that macro fusion does \emph{not} impact the reliability of 
\attack as a deterministic attack primitive.
That is, we consistently observed in all of our attacks that macro fusion depends solely on the 
architectural program state, \ie opcode types and their alignments, and hence a given code path 
always results in the same, deterministic number of interrupts.

To the best of our knowledge, \attack contributes the first
methodology to independently research and reverse-engineer macro fusion
optimizations in Intel processors.
While our observations confirm that macro fusion behaves as specified,
we consider a precise understanding of macro fusion of particular importance for
compile-time hardening techniques that balance conditional code paths (\S\ref{sec:countermeasure}).

\subsection{Instruction-Level Page Access Traces}

\para{Leakage Model}
\attack complements the coarse-grained 4\,KiB spatial resolution of previous
page fault-driven attacks with a fully deterministic temporal dimension.
By interrupting after every instruction and querying page-table
\enquote{accessed} bits, \attack adversaries obtain an instruction-granular
trace of page visits performed by the enclave.
This trace may reveal private branch decisions whenever a secret-dependent execution
path does not access the exact same set of code and data pages 
at every instruction offset in both branches.
Importantly, even when both execution paths access the same sequence of code and
data pages, and hence remain indistinguishable for a traditional page-fault
adversary~\cite{xu2015controlled}, we show below that compilers may in practice
still emit unbalanced instruction counts between page accesses in both branches. 
\Cref{sec:limit,sec:countermeasure} elaborate further on the limitations of this
leakage model and the precise requirements for static code balancing solutions.

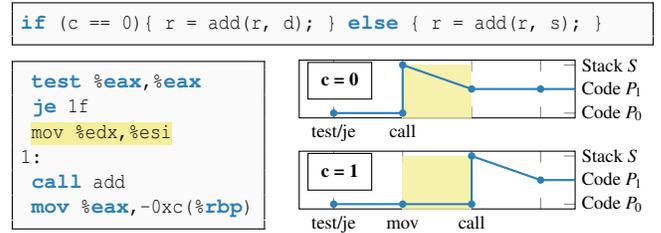
\begin{figure}[t]
  
\begin{lstlisting}
if (c == 0){ r = add(r, d); } else { r = add(r, s); }
\end{lstlisting}
  \vspace{-.2em}  
  {
    \centering
  \begin{minipage}{.39\linewidth}
    \lstset{language=[x64]Assembler}
\begin{lstlisting}
 test %eax,%eax
 je 1f
 @\hl{mov \%edx,\%esi}@
1:
 call add
 mov %eax,-0xc(%rbp)
\end{lstlisting}
  \end{minipage}\hfill
    \hspace{.5cm}
  \begin{minipage}{.6\linewidth}
    \vspace{-.5em}
    \scriptsize
\begin{tikzpicture}
 \pgfplotstableread[col sep=comma]{if.csv}{\Data}
 \pgfplotstableread[col sep=comma]{else.csv}{\DataTwo}
 \begin{groupplot}[
     group style = {group size = 1 by 2, vertical sep=4ex},
     height = 2.35cm,
     width = 1.033\linewidth,
     legend style={empty legend,font=\bfseries},
     ytick=data,
     ytick pos = right,
     symbolic y coords={Code $P_0$,Code $P_1$,Stack $S$},
     xtick = {0,1,2,3,4},
     xticklabel style={text height=1.5ex},
     xticklabels = none,
     xmin=-0.5,
     xmax=3.5,
    ]
    \nextgroupplot[legend pos=north west, xticklabels from table = {\Data}{Inst}]
      \addlegendentry{c = 0}
        \draw[fill=yellow, opacity=0.8,draw=none] (axis cs: 1,[normalized]-1) rectangle (axis cs: 2,[normalized]2);
      \addplot[blue,thick,mark options={draw=blue,fill=blue},mark=*,mark size=1] table [x = IRQ, y = Page] {\Data};

    \nextgroupplot[legend pos=north west, xticklabels from table = {\DataTwo}{Inst}] 
      \addlegendentry{c = 1}
        \draw[fill=yellow, opacity=0.8,draw=none] (axis cs: 1,[normalized]-1) rectangle (axis cs: 2,[normalized]2);
      \addplot[blue,thick,mark options={draw=blue,fill=blue},mark=*,mark size=1] table [x = IRQ, y = Page] {\DataTwo};
  \end{groupplot}
\end{tikzpicture}

  \end{minipage}
  }
    \caption{Balanced if/else statement (top), compiled to assembly (left).
             Precise page-aligned, intra-cache line conditional control flow can be deterministically reconstructed with instruction-granular \attack page access traces (right).  }
    \label{fig:ifelse}
\end{figure}

\para{If/Else Statement}
Conditional branches are pervasive in all applications~\cite{xu2015controlled,hahnel2017high,lee2017inferring,hosseinzadeh2018mitigating}, 
but even side-channel hardened cryptographic software may assume that carefully aligned if/else statements or tight loops 
cannot be reliably reconstructed (\S\ref{sec:wolfssl}).
\Cref{fig:ifelse} provides a minimal example of an if statement that has been hardened 
using a balancing else branch, \eg as in the Montgomery Ladder algorithm.
The corresponding assembly code, as compiled by \texttt{gcc}, indeed only differs in a single x86 instruction 
that can fit entirely within the same page and cache line.
This if branch is hence indistinguishable for a page-fault or cache adversary.
While finer-grained, branch prediction side channels may still be able to reconstruct the branch outcome, 
these attacks typically require several runs of the victim and can be trivially addressed by flushing the branch predictor on an enclave exit.

\Cref{fig:ifelse} illustrates how \attack can {deterministically} reconstruct the branch outcome 
merely by counting the number of instructions executed on the $P_0$ code page containing the if branch 
before control flow is eventually transferred to the $P_1$ code page containing the \texttt{add} function, 
as revealed by probing the \enquote{accessed} bit in the corresponding page-table entry.
The example furthermore highlights that even if all of the code were to fit on a
single code page $P_0=P_1$, \attack adversaries could still distinguish both branches by
comparing the relative position of the data access to the stack page $S$
performed by the \texttt{call} instruction.
In particular, while traditional page-fault adversaries always see the same \emph{page fault sequence} 
$(P_0,S,P_1)$, independent of the secret, \attack enriches this information with precise instruction counts, 
resulting in distinguishable \emph{instruction-level page access traces} $(P_0,P_0,S,P_1)$ vs.\ $(P_0,P_0,P_0,S,P_1)$.

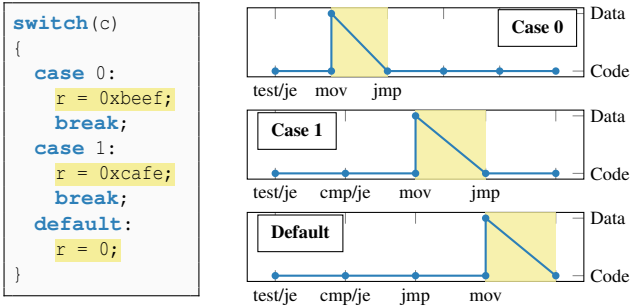
\begin{figure}[t]
  \begin{minipage}{.3\hsize}
\begin{lstlisting}
switch(c)
{
  case 0:
    @\hl{r = 0xbeef;}@
    break;
  case 1:
    @\hl{r = 0xcafe;}@
    break;
  default:
    @\hl{r = 0;}@
}
\end{lstlisting}
  \end{minipage}
  \hfill
  \begin{minipage}{.65\hsize}
  \scriptsize
\begin{tikzpicture}
 \pgfplotstableread[col sep=comma]{data.csv}{\Data}
 \pgfplotstableread[col sep=comma]{data2.csv}{\DataTwo}
 \pgfplotstableread[col sep=comma]{data3.csv}{\DataThree}
 \begin{groupplot}[
     group style = {group size = 1 by 3, vertical sep=4ex},
     height = 2.5cm,
     width = 1.1\linewidth,
     legend style={empty legend,font=\bfseries},
     ytick=data,
     ytick pos = right,
     symbolic y coords={Code,Data},
     xtick = {0,...,4},
     xticklabel style={text height=1.5ex},
     xticklabels = none,
    ]
    \nextgroupplot[xticklabels from table = {\Data}{Inst}]
      \addlegendentry{Case 0}
        \draw[fill=yellow, opacity=0.8,draw=none] (axis cs: 1,[normalized]-1) rectangle (axis cs: 2,[normalized]2);
      \addplot[blue,thick,mark options={draw=blue,fill=blue},mark=*,mark size=1] table [x = IRQ, y = Page] {\Data};

    \nextgroupplot[legend pos=north west,xticklabels from table = {\DataTwo}{Inst}]
      \addlegendentry{Case 1}
        \draw[fill=yellow, opacity=0.8,draw=none] (axis cs: 2,[normalized]-1) rectangle (axis cs: 3,[normalized]2);
      \addplot[blue,thick,mark options={draw=blue,fill=blue},mark=*,mark size=1] table [x = IRQ, y = Page] {\DataTwo};

    \nextgroupplot[legend pos=north west, xticklabels from table = {\DataThree}{Inst}] 
      \addlegendentry{Default}
        \draw[fill=yellow, opacity=0.8,draw=none] (axis cs: 3,[normalized]-1) rectangle (axis cs: 4,[normalized]2);
      \addplot[blue,thick,mark options={draw=blue,fill=blue},mark=*,mark size=1] table [x = IRQ, y = Page] {\DataThree};
  \end{groupplot}
\end{tikzpicture}

  \end{minipage}
  \hspace{-.5cm}
  \caption{
      Conditional data assignments in a page-aligned switch statement (left) deterministically leak through their relative positions in the precise, instruction-granular page access traces extracted by \attack (right).
  }
  \label{fig:switch}
\end{figure}

\para{Switch-Case Statement}
As a further example, \cref{fig:switch} illustrates precise control-flow
recovery in a switch-case statement, where the code blocks again fall entirely
within a single page and cache line, and where the same data is accessed in every case.
While traditional page-fault adversaries always observe an identical, input-independent access sequence to the code and data pages,
and the tight sequence of conditional jumps poses a considerable challenge for branch prediction adversaries~\cite{lee2017inferring},
\attack deterministically reveals the entire control flow through the \emph{relative} position of the data access in the instruction-granular page access traces.

\subsection{Defeating Branch Shadowing Defenses}
\label{sec:zigzag}
To highlight the importance of \attack for non-cryptographic applications, we employ its improved 
resolution to defeat a state-of-the-art compiler defense~\cite{hosseinzadeh2018mitigating} against branch predictor leakage.
This again shows that \attack changes the attack landscape and requires orthogonal 
mitigations when compared to microarchitectural side channels.

\begin{figure}
    \centering
    \includegraphics[width=.9\linewidth]{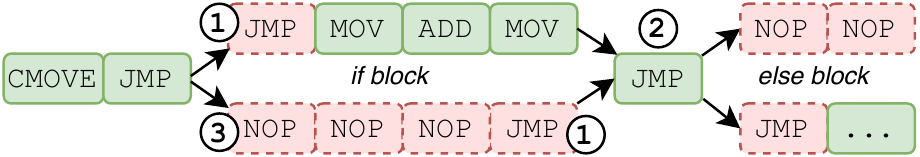} 
    \caption{
        Compiler mitigation~\cite{hosseinzadeh2018mitigating} for branch prediction side channels.
        \attack reveals control flow via the number of instructions executed on the trampoline page (red, dashed).
}
    \label{fig:zigzag}
\end{figure}

\para{Branch Shadowing Mitigation}
Lee \etal\cite{lee2017inferring} first proposed Zigzagger, an automated compile-time 
approach to defend against branch-shadowing attacks by rewriting conditional branches 
using \texttt{cmov} and a tight trampoline sequence of unconditional jump instructions.
However, the security of their compiler transformation critically relies on the trampoline 
sequences being non-interruptible, and several proof-of-concept attacks on Zigzagger have 
been demonstrated using precise interrupt capabilities~\cite{vanbulck2017SGXStep,van2018nemesis,gyselinck2018off}.
In response, Hosseinzadeh \etal\cite{hosseinzadeh2018mitigating} designed an improved compiler 
mitigation that employs runtime randomization to dynamically shuffle jump blocks on the trampoline area, 
thereby effectively hiding branch targets and making branch shadowing attacks probabilistically infeasible.
\Cref{fig:zigzag} illustrates how conditional branches are redirected through randomized jump 
locations \textcircled{\footnotesize 1} on the trampoline page, while ensuring that all 
jumps \textcircled{\footnotesize 2} outside of the trampoline are always executed in the same order.
Finally, to protect against timing attacks, trampoline code is explicitly balanced with dummy 
instructions \textcircled{\footnotesize 3} to compensate for skipped blocks in the instrumented code.

\para{Case-Study Attack}
We evaluated \attack on the open-source\footnote{Branch shadowing mitigation: \url{https://github.com/SSGAalto/sgx-branch-shadowing-mitigation}}
release of the compiler hardening scheme~\cite{hosseinzadeh2018mitigating} based on LLVM 6.0.
First, we found that the dummy instruction balancing pass is not always entirely accurate and 
may result in execution paths that differ slightly by one or two instructions (\cf \cref{app:zigzag}).
Crucially, while such subtle deviations would indeed very likely not be exploitable through timing, 
as originally envisioned by the mitigation, we experimentally validated that the unbalanced paths 
can be fully deterministically distinguished by \attack{} adversaries.
Second, even when the code paths are perfectly balanced, 
\cref{fig:zigzag} illustrates that merely counting the \emph{number} of instructions executed 
on the trampoline page deterministically reveals whether the victim is executing balancing dummy 
code in a trampoline block or the actual if block on the instrumented code page.
Note that the compiler carefully maintains a constant jump order when moving back and forth between the trampoline 
area and the instrumented code, ensuring that the execution remains oblivious to classical page-fault 
adversaries~\cite{xu2015controlled,shinde2016preventing} who will always observe the exact same \emph{sequence} 
of pages regardless of the actual code blocks being executed.

\section{Unleashing \attack on WolfSSL} \label{sec:wolfssl}
WolfSSL is a prominent, FIPS-certified solution officially supporting 
Intel SGX~\cite{wolfsslfipssgx}. 
In a case study on the WolfSSL cryptographic library, 
we show that \attack enables attacks that were not previously possible without 
a deterministic and fine-grained leakage model.
In \cref{sec:wolfbeea}, we outline our 
controlled-channel attack using \attack to precisely 
recover the full execution trace of WolfSSL's implementation of the binary extended Euclidean algorithm (BEEA),
which is used for modular inversion of cryptographic secrets in DSA, ECDSA, and RSA.
Precise recovery of the full execution flow of BEEA enables new single-trace algorithmic 
attacks on both DSA signing and RSA key generation, as demonstrated in \cref{sec:wolfdsa,sec:wolfrsa}, respectively.
Finally, we apply \attack to bypass incomplete side-channel mitigations and recover deterministic partial information on ECDSA signatures, which allows for efficient key recovery via lattices. 

\para{Experimental Setup}
Our experimental setup includes a desktop Intel Core i7-7700 CPU that supports Intel SGX and is 
updated with the latest microcode (0xca) running Ubuntu 16.04 with kernel 4.14.0-72-generic.
We use the SGX-Step~\cite{vanbulck2017SGXStep} framework v1.4.0 to implement our attacks on the latest stable WolfSSL version 4.2.0. 
WolfSSL officially supports compilation for Intel SGX enclaves.
We implemented our key recovery attacks in SageMath version 8.8.

\subsection{\attack on BEEA} \label{sec:wolfbeea}
Computing the modular inverse or greatest common divisor (GCD) using the binary extended Euclidean algorithm (BEEA)
has previously exposed cryptographic implementations to side-channel attacks~\cite{aciiccmez2007power,garcia2017constant,weiser2018single}.
The BEEA, as shown in \cref{alg:beea}, is not constant time and can leak various bits of its input. 
However, previous attacks are limited to recovering only partial and noisy information about the secret input. 
This limitation stems from low spatial resolution and the presence of noise.
For instance, a cache- or page-level attacker who can distinguish which arithmetic subroutines have 
been invoked cannot determine the outcome of the comparison at line 13 since both directions 
of the branch generate exactly the same sequence of memory access patterns.
In addition, the arithmetic functions may fit within the same page and become indistinguishable for a page-level adversary.
Alternatively, a cache attacker may try to track the outcome of these branches within the
same page by tracking the corresponding instruction cache lines for the BEEA subroutine. 
However, a compact implementation of this algorithm can fit multiple branches within the same cache line. 
While some microarchitectural attacks on the instruction stream may leak some of these low-level branch outcomes, 
they are all prone to various amounts of noise~\cite{aldaya2018port,van2018nemesis,lee2017inferring,evtyushkin2018branchscope,huo2020bluethunder}.

\begin{algorithm}[t]
  \caption{Modular inversion using the BEEA. In the optimized compact implementation when the modulus is odd, highlighted (blue) statements are removed.}
  \label{alg:beea}
  \begin{algorithmic}[1]
    \Procedure{modInv}{$u$, modulus $v$}
    \State $b_i \gets 0$ $d_i \gets 1, u_i \gets u, v_i = v$, \HiLi[1.8cm]{$a_i \gets 1$, $c_i \gets 0$} 
    \While{$isEven(u_i)$}
      \State $u_i \gets u_i / 2$
      \If {$isOdd(b_i)$} 
      \State $b_i \gets b_i - u$, \HiLi[1.5cm]{$a_i \gets a_i + v$}
      \EndIf
      \State $b_i \gets b_i / 2$, \HiLi[1.5cm]{$a_i \gets a_i / 2$}
    \EndWhile

    \While{$isEven(v_i)$}
      \State $v_i \gets v_i / 2$
      \If {$isOdd(d_i)$} 
      \State $d_i \gets d_i - u$, \HiLi[1.5cm]{$c_i \gets c_i + v$} 
      \EndIf
      \State $d_i \gets d_i / 2$, \HiLi[1.5cm]{$c_i \gets c_i / 2$} 
    \EndWhile

    \If{$u_i > v_i$}
      \State $u_i \gets u_i - v_i$, $b_i \gets b_i - d_i$, \HiLi[1.5cm]{$a_i \gets a_i - c_i$}
    \Else
      \State $v_i \gets v_i - u_i$, $d_i \gets d_i - b_i$, \HiLi[1.5cm]{$c_i \gets c_i - a_i$}
      \State 
    \EndIf
    \Return $d_i$
    \EndProcedure
  \end{algorithmic}
\end{algorithm}

WolfSSL supports two different BEEA implementations in subroutines 
\texttt{fp\_invmod\_slow} and \texttt{fp\_invmod}.\footnote{
  \texttt{fp\_invmod\_slow} and \texttt{fp\_invmod} can be found at line 885 and 1015 of \url{https://github.com/wolfSSL/wolfssl/blob/48c4b2fedc/wolfcrypt/src/tfm.c}, respectively.
}
The former is a straightforward implementation, 
and the latter is a compact implementation that only supports odd moduli. 
We analyze both implementations and show how to use \attack to recover the runtime 
control flow of these implementations deterministically and without noise.

\para{Binary Layout of Modular Inversion}
After compilation, the subroutines \texttt{fp\_iseven} and \texttt{fp\_isodd} 
are simply inlined within the same page as their caller \texttt{fp\_invmod\_slow}.
However, the arithmetic functions \textit{A}=\texttt{fp\_add}, \textit{C}=\texttt{fp\_cmp}, 
\textit{D}=\texttt{fp\_div\_2}, and \textit{S}=\texttt{fp\_sub} are external calls and reside in a new page. 
Analyzing these arithmetic functions (A, C, D, S), including their internal subroutines, shows that they span $2,895$ bytes. Hence, it is reasonable to assume 
that they can fit into a single 4\,KiB page, thus preventing a page-level attacker from 
distinguishing them at runtime altogether.
In addition, even assuming they do not align within the same page, 
reconstructing the exact execution flow is still impossible.  For example, the transition from \textit{S} to \textit{D} 
can result from multiple different code paths. 
The instructions for \texttt{fp\_invmod\_slow} can fit into fewer than 6 cache lines with multiple 
basic blocks\footnote{A basic block is a code sequence that has no branches in and out.} overlapping within the same line.

WolfSSL also supports a modified version of BEEA, \texttt{fp\_invmod} specialized to the 
case of odd modulus, which is used for RSA $q^{-1} \bmod p$ (\S\ref{sec:wolfrsa}) and DSA $k^{-1} \bmod n$ 
(\S\ref{sec:wolfdsa}). The control flow and overall layout for \texttt{fp\_invmod} are
similar to the above implementation but it is more compact, as some of the arithmetic 
statements have been removed. 
\texttt{fp\_invmod} can fit into fewer than 4 cache lines with multiple overlapping basic blocks. 

\begin{figure}[tb]
 \centering
 \includegraphics[width=.89\linewidth]{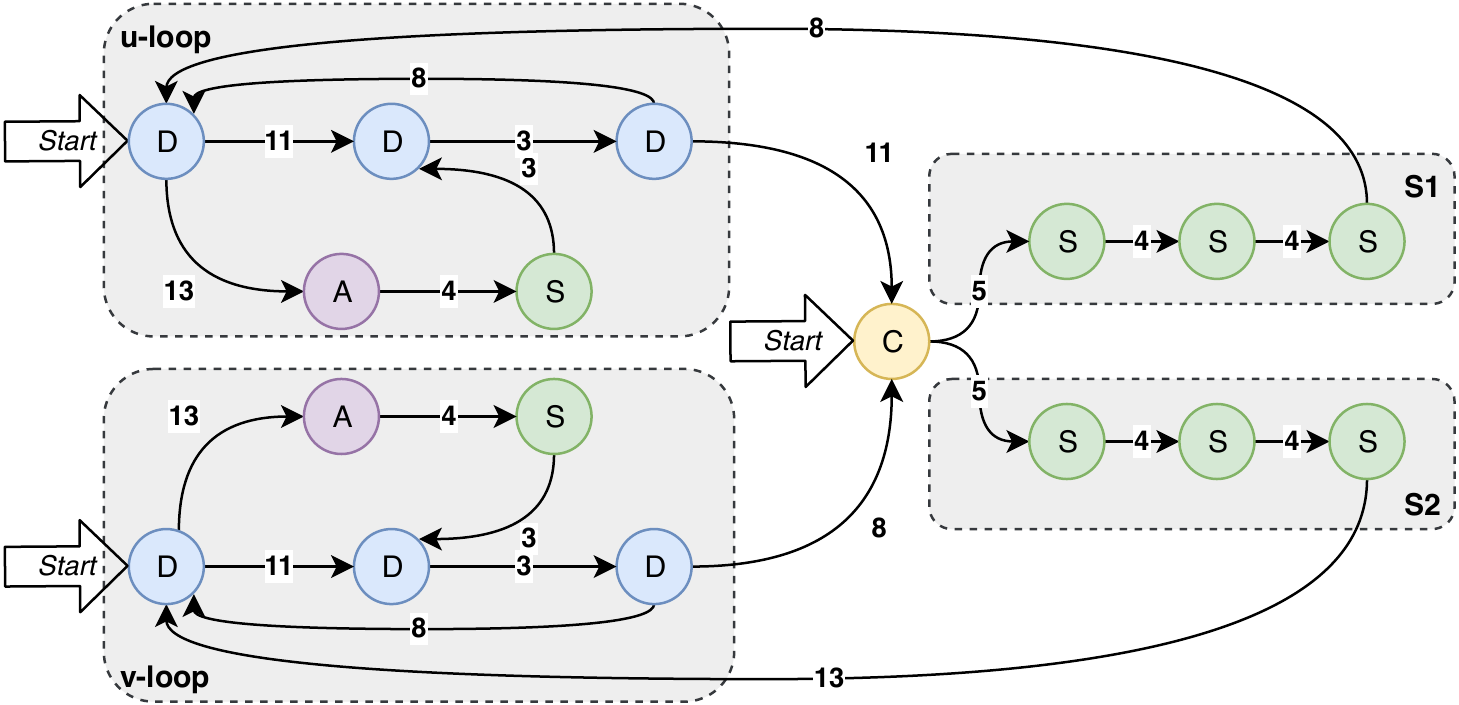} 
 \caption{Control flow of the BEEA as implemented by \texttt{fp\_invmod\_slow}. 
 Each circle (D=div, C=cmp, S=sub, A=add) represents a call to a function in the page that holds 
 these arithmetic functions. We count the exact number of instructions between two consecutive 
 invocations that hit this page. The instruction counts reveal branch outcomes.}
 \label{fig:fp_invmod_slow}
 \end{figure}

\para{Recovering BEEA Control-Flow Transfers}
We analyzed the runtime control flow of \texttt{fp\_invmod\_slow} by matching 
its disassembly with the execution trace we recovered from running \attack.
\cref{fig:fp_invmod_slow} shows the control flow transfers at page-level granularity 
for the page corresponding to \texttt{fp\_invmod\_slow} and the page corresponding to
arithmetic functions (Circles). Additionally, the weight of each arrow shows the number 
of instructions that are executed for \texttt{fp\_invmod\_slow} before accessing 
the page corresponding to arithmetic functions. The division loop for $u_i$ (\textit{u-loop}) 
and $v_i$ (\textit{v-loop}) have a similar control flow. In addition, the two blocks of substitutions 
after the comparison of $u > v$ have similar control flow for both the left \textit{S1} 
and right \textit{S2} direction. Only certain transitions are viable from these blocks 
to division loops during the computation of the modular inverse. For example, \textit{S2} always 
goes to \textit{v-loop} and \textit{S1} always goes to \textit{u-loop}. 
Since these instruction counts are distinguishable for transitions that are 
related to conditional statements, we can use a trace consisting of a vector of 
these weights in the graph to infer the outcome of the conditional statement. 

\begin{figure}[tb]
\centering
\includegraphics[width=1\linewidth]{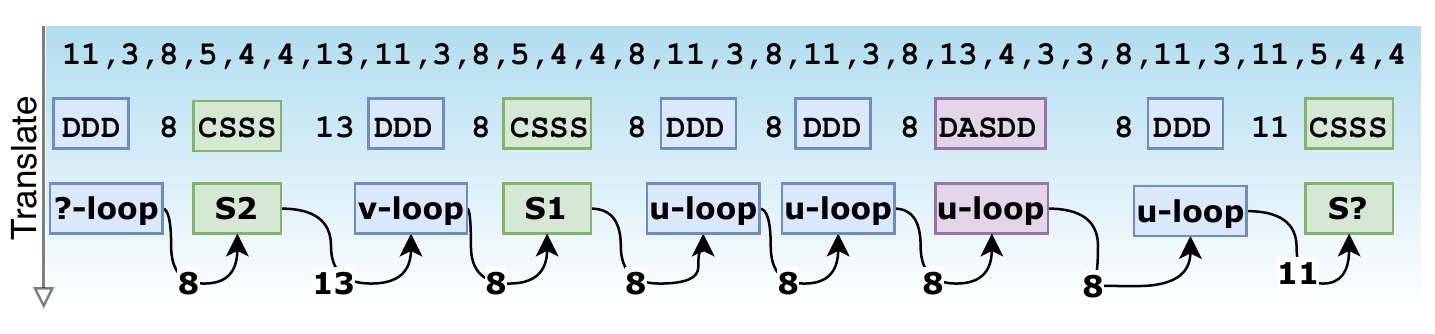} 
\caption{An example cut of a trace that is recovered from \texttt{fp\_invmod\_slow}.
First, the weights are replaced according to Rules 1, 2, and 3. Then other 
transitions (Rules 4 and 5) are used to recover the whole control flow sequence.
}
\label{fig:example_trace}
\end{figure}
  
With a trace including the weights of instruction counts collected between two consecutive 
accesses to the page that holds the arithmetic operations (A, C, D, S), 
we apply a set of divide-and-conquer rules to reconstruct the control flow for \texttt{fp\_invmod\_slow}. 
These rules start by translating the recovered weights to corresponding generic blocks. 
For example, every time the algorithm executes an iteration of a division loop (u/v-loop), 
we observe either the sequence $D \rightarrow D \rightarrow D$, 
or the sequence $D \rightarrow A \rightarrow S \rightarrow D\rightarrow D$. 
Each of these sequences generates a consistent set of weights. 
Similarly, \textit{S1} or \textit{S2} always generates a sequence like $C \rightarrow S \rightarrow S \rightarrow S$. 
After translating these generic blocks, we can use the remaining transitions 
to distinguish the exact blocks, i.e., we can recover whether a \textit{S1} or \textit{S2} 
followed by a set of division loops is equal to a transition from \textit{S1} to \textit{u-loop} or 
transition from \textit{S2} to \textit{v-loop}. These rules are summarized as follow:

\begin{CompactItemize}
  \item \textbf{Rule 1:} $? \xrightarrow[]{\text{11}} ? \xrightarrow[]{\text{3}} ? = D \rightarrow D \rightarrow D$. 
  \item \textbf{Rule 2:} $? \xrightarrow[]{\text{13}} ? \xrightarrow[]{\text{4}} ? \xrightarrow[]{\text{3}} ?\xrightarrow[]{\text{3}} ? = D \rightarrow A \rightarrow S \rightarrow D \rightarrow D$. 
  \item \textbf{Rule 3:} $? \xrightarrow[]{\text{5}} ? \xrightarrow[]{\text{4}} ? \xrightarrow[]{\text{4}} ? = C \rightarrow S \rightarrow S \rightarrow S$. 
  \item \textbf{Rule 4:} $S? \xrightarrow[]{\text{13}} ? = S2 \rightarrow$ \textit{v-loop}.
  \item \textbf{Rule 5:} $S? \xrightarrow[]{\text{8}} ? = S1 \rightarrow$ \textit{u-loop}.
\end{CompactItemize}

We first replace some of the weights according to Rules 1, 2, and 3, which identify if we are in 
a division loop (u-loop or v-loop) or a comparison and substitution block (S?). 
Then based on the other transitions (Rule 4 and 5), 
we can determine which state of the comparison and substitution block we have moved from, and which division loop 
we have moved to within the trace. An example sequence from the execution of \textit{fp\_invmod\_slow} 
and its translation to the control flow transitions is given in \cref{fig:example_trace}.

\begin{figure}[tb]
  \centering
  \includegraphics[width=.73\linewidth]{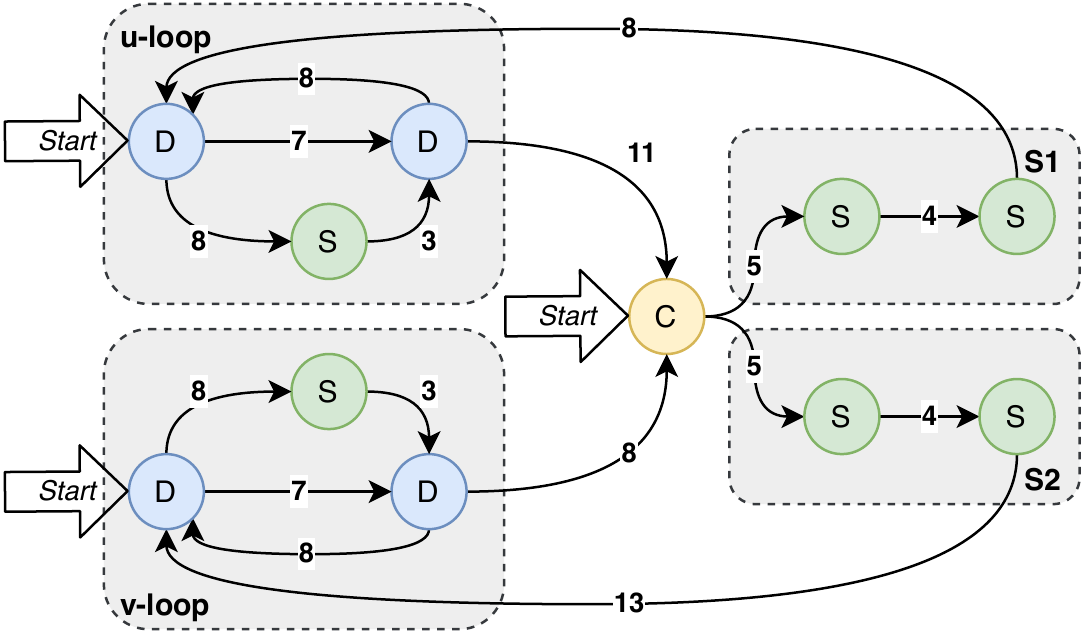} 
   \caption{Control flow of BEEA in \texttt{fp\_invmod}.}
   \label{fig:fp_invmod}
 \end{figure}

For the compact implementation in \textit{fp\_invmod}, we apply the same approach. 
\cref{fig:fp_invmod} shows the control flow for this implementation after runtime analysis using \attack. 
Similarly, we define a set of rules to translate the trace of instruction counts to control 
flow transfers of BEEA. Based on \cref{fig:fp_invmod}, we modify the first three rules 
as follows to support control-flow recovery based on the same approach:\footnote{{Rule 4} and {5} remain the same.}

\begin{CompactItemize}
  \item \textbf{Rule 1:} $? \xrightarrow[]{\text{7}} ? = D \rightarrow D$. 
  \item \textbf{Rule 2:} $? \xrightarrow[]{\text{8}} ? \xrightarrow[]{\text{3}} ?\xrightarrow[]{\text{3}} ? = D  \rightarrow S \rightarrow D$. 
  \item \textbf{Rule 3:} $? \xrightarrow[]{\text{5}} ? \xrightarrow[]{\text{4}} ? = C \rightarrow S \rightarrow S $. 
\end{CompactItemize}

\subsection{Single-Trace Attack on DSA Signing} \label{sec:wolfdsa}
In contrast to previous attacks on BEEA that leak partial information 
about the nonce~\cite{garcia2017constant}, \attack\  
recovers virtually the entire control flow from the execution of this 
implementation with 100 percent precision. 
As a result, we can perform a single-trace attack on the DSA signing operation. 
In \Cref{sec:crypto}, we generalize this attack and expose 
multiple vulnerabilities in the \textit{Libgcrypt} library.

\para{DSA Key Recovery}
WolfSSL uses \texttt{fp\_invmod} to compute the modular inversion 
of $k_{inv} = k^{-1} \bmod n$, where $n$ is an odd prime. 
Since we can recover the exact control flow of this computation and the modulus $n$ is public, 
we simply step through the execution trace of \cref{alg:beea}, 
applying each step of the computation according to the recovered trace to compute $k_{inv}$ bit by bit. 
After recovering $k_{inv}$, recovering the full nonce and private key 
is trivial: $k = k_{inv}^{-1} \bmod n$, $x = r^{-1}(sk - h) \bmod n$.

\para{Evaluation}
To attack 160-bit DSA, we used a combination of pages in a page-level 
controlled-channel attack to first reach the beginning of the modular 
inversion operation for DSA. Then we start \attack over the code page for \texttt{fp\_invmod}.
We executed this attack for 100 different signing operations.
On average, this attack issues 22,000 IRQs and takes 75 ms to iterate over an 
average of 6,320 steps for each signature generation. 
Out of 100 experiments, our single-trace attack successfully recovered the 
full control flow and the key using the algorithm above,
implying that \attack reliably reconstructs the entire execution flow.
As a result, a single-trace attack on DSA can be executed without 
the need for multiple signatures. 

\subsection{Single-Trace Attacks on RSA KeyGen} \label{sec:wolfrsa}
During RSA key generation, WolfSSL checks if a potential prime $p$ 
is coprime with $e$ by checking if $gcd(e, p-1)$ is equal to $1$. 
This step uses the textbook greatest common divisor (GCD) algorithm, 
which simply performs a series of divisions.  This algorithm appears to be less vulnerable
to control-flow-based key recovery.
However, in a later stage, WolfSSL computes $d = e^{-1} \bmod \lambda(N)$ 
and the CRT parameter $q^{-1} \bmod p$ using the BEEA.
WolfSSL always generates the CRT parameters during RSA key generation.\footnote{
  \texttt{wc\_MakeRsaKey} at \url{https://github.com/wolfSSL/wolfssl/blob/48c4b2fedc/wolfcrypt/src/rsa.c\#L3726} invokes BEEA multiple times during RSA Key generation.
}

\para{Key Recovery from a $q^{-1} \bmod p$ Trace}
Compared to  $k^{-1} \bmod n$, this attack is more challenging 
since in this case, both operands $p$ and $q$ are unknown.
We give a novel and efficient attack that recovers the private RSA parameters $p$ and $q$ using \attack.
We use the relationship of the public modulus $N = pq$ and the execution 
trace of the BEEA on $q^{-1} \bmod p$, which provides enough information to 
recover the factorization of $N$. The main idea is that the BEEA algorithm works sequentially from the least significant bits of $p$ and $q$.  Thus if we iteratively guess bits of $p$ and $q$ starting from the least significant bits, we can verify that a guess matches the relevant steps of the BEEA execution trace, as well as the constraint that $N = pq$ for the bits guessed so far, and eliminate guesses that do not.  This algorithm resembles the branch-and-prune 
algorithm of~\cite{heninger2009reconstructing}, with new constraints.

We propose \cref{alg:branchprune} to recover $p$ and $q$ using only knowledge of $N$ 
and the execution trace of the BEEA on $q^{-1} \bmod p$. 
The algorithm starts by initializing a list of hypotheses for values of the least significant bits of $q$ and $p$.
Each hypothesis keeps track of the current step, bit position $b$, and the hypothesized 
values of $p_s=p \bmod 2^b$ and $q_s=q \bmod 2^b$. Among the four possible assignments 
for the $(b+1)^{st}$ bits of $p$ and $q$ in Step 7, there will be 
two choices satisfying the constraint that $pq \equiv N \bmod{2^{b+1}}$. 
For these new guesses, we evaluate the BEEA algorithm up to the number of bits 
guessed so far, and check this deterministic algorithm evaluation on the guess 
against the ground truth execution trace $t$. We then do a depth-first search prioritized by the number of steps in which the algorithm executed correctly, and terminate when we have found a candidate for which $pq = N$ holds. 

\begin{algorithm}[t]
  \caption{Recovering $p$ and $q$ from trace of $q^{-1} \bmod p$.}
  \label{alg:branchprune}
  \begin{algorithmic}[1]
    \Procedure{recover\_pq}{trace $t$, modulus $N$}
    \State $h  \gets (-test\_t(t,1,1),1,1,1)$
    \While{$h$}
      \State $steps,b,p,q \gets hpop(h)$
      \If{$p.q = N$}
        \Return $p, q$ 
      \EndIf
      \State $g \gets (p,q),(p+2^b,q),(p,q+2^b),(p+2^b,q+2^b)$
      \For{$p_{s},q_{s}$ in $g$}
        \If {$mod(p_{s}.q_{s},2^{b+1}) = mod(N,2^{b+1})$}
          \State $hpush(h,(-test\_t(trace,p_{s},q_{s}),b+1,p_{s},q_{s}))$
        \EndIf
      \EndFor
    \EndWhile 
    \EndProcedure
  \end{algorithmic}
\end{algorithm}

\para{Evaluation}
We executed an attack similar to \cref{sec:wolfdsa} to collect traces from the modular 
inversion of $q^{-1} \bmod p$, as it is computed by \texttt{fp\_invmod\_slow}. 
We tried this attack on 100 different 2048-bit RSA key generations. 
On average, we iterate over 39,400 steps by issuing 106490 IRQs in 365 ms. 
However, the average time to collect a trace can take up to a second depending on how 
fast the prime numbers are chosen. The attack takes 20 seconds to recover the key from a trace.
All 100 trials of the attack successfully recovered the keys.

\para{Key Recovery from an $e^{-1} \bmod \lambda(N)$ Trace}
In contrast to previous attacks on this computation~\cite{cabrera2017side}, 
we propose a different algorithmic attack that takes advantage of the fact that \attack\ can recover 
the entire control flow of this algorithm. As a result, a single-trace attack can be carried out for any value of $e$, both large or small. This shows that the proposed masking countermeasure in~\cite{cabrera2017side}
is insecure against our strong \attack adversary.

Our goal is to recover the RSA primes $p$ and $q$ using the trace of the BEEA 
for $d = e^{-1} \bmod \lambda(N)$. The modulus $N$ and the public exponent $e$ are known, 
while $\lambda(N)$ is secret. We present a modified branch-and-prune technique in \cref{alg:branchprunelambda} 
that recovers the factors $p$ and $q$ for a large fraction of generated RSA keys. 

The main idea is to iteratively guess bits of $p$ and $q$ starting from the least significant bits, then verify that $pq = N$ and the relevant steps of the BEEA execution trace match the guess so far.  However, the BEEA is computed on $e$ and $\lambda(N) = (p-1)(q-1)/\gcd(p-1,q-1)$.  We do not know $\gcd(p-1,q-1)$ and must guess it for this algorithm, but with high probability it only has small factors and can be brute forced.  For simplicity, we specialize to the case of $\gcd(p-1,q-1) = 2^i$ for small integer $i$ below, but the analysis can be extended to other candidate small primes with more brute force effort.

For each guess $2^i$ for $\gcd(p-1,q-1)$, we iteratively generate guesses for $p_s$ and $q_s$, compute $\phi_s = (p_s-1)(q_s-1)$ and then $\lambda_s = \phi_s/2^i$. We compare the execution trace $t$ to the execution trace for $\lambda_s$ and $e$. The algorithm either returns $p$ and $q$ 
or it fails to recover $p$ and $q$ if $\phi/\lambda(N) \neq 2^i$.

\begin{algorithm}[t]
  \caption{Recovering $p$ and $q$ from trace of $e^{-1} \bmod \lambda$.}
  \label{alg:branchprunelambda}
  \begin{algorithmic}[1]
    \Procedure{recover\_pq}{trace $t$, $e$, modulus $N$}
    \State $h  \gets (-test\_t(t,0,e),1,1,1)$
    \While{$h$}
      \State $steps,b,p,q \gets hpop(h)$
      \If{$p.q = N$}
        \Return $p, q$ 
      \EndIf
      \State $g \gets (p,q),(p+2^b,q),(p,q+2^b),(p+2^b,q+2^b)$
      \For{$p_{s},q_{s}$ in $g$}
        \If {$mod(p_{s}.q_{s},2^{b+1}) = mod(N,2^{b+1})$}      
      	\State $\phi=(p_s-1)(q_s-1)$
        \For{$i=1,\ldots,2^\ell$}
	      	\If {$p_sq_s>N$ or $mod(\phi, 2^i)\neq 0$}
        	    \State continue
    	    \EndIf
	        \State $\lambda=\phi/2^i$
	        \State $newsteps = test\_t\_lamda(t,\lambda,e)$
               \If{$newsteps >= b+1$}:
                  \State $hpush(h,(-newsteps,b+1,p_s,q_s))$
               \EndIf
	        \EndFor
        \EndIf
      \EndFor
    \EndWhile 
    \Return \texttt{fail}
    \EndProcedure
  \end{algorithmic}
\end{algorithm}

\para{Analysis}
The algorithm will succeed whenever $\phi/\lambda = 2^i$ for small $i$. 
For non-powers of $2$ the test against the BEEA execution trace in Step 15 will likely fail, and cause this branch to be pruned. Since $p=2p'+1$ and $q=2q'+1$ for 
some $p', q' \in \Z$, we have $\lambda(N)= \lcm(p-1,q-1) = 2 \lcm(p',q')$. From the prime number theorem~\cite{selberg1949elementary}, the probability that two random integers are coprime 
converges to $\prod_{p \in primes}(1-1/p^2) = \frac{6}{\pi^2} \approx 61\%$ as the size of the integers increases.
In other words, if we run~\cref{alg:branchprunelambda} for only $i=1$, it will succeed $61\%$ of 
the time when $p'$ and $q'$ are actually coprime. If we allow $p'$ and $q'$ to have even factors 
we obtain a probability of $\prod_{p \in primes, p>2}(1-1/p^2) = \frac{8}{\pi^2} \approx 81\%$. 
This means that even for a modest number of iterations, e.g. $\ell=8$, we have nearly $81\%$ success probability. These estimates are confirmed by our experiments.

\para{Evaluation}
We tried this attack on 100 different key generation efforts (2048-bit key). 
On average, we iterate over 81,090 steps by issuing 230,050 interrupts per attack in 800ms. 
The average time to collect a trace is about a second and the attack takes about 30 seconds 
to successfully recover the key for 81\% of the keys when $\lcm((p-1)(q-1)) \equiv (p-1)(q-1)/2^{i}$. 

\para{Revisiting Masking Protection}
Earlier attacks required brute forcing over $e$~\cite{weiser2018single}. Our algorithm works for arbitrary, even full length $e$. Thus increasing the size of $e$ by choosing a bigger public exponent 
or masking is not sufficient to mitigate our attack. 
Aldaya \etal\cite{aldaya2017spa} proposed masking $e$ by computing $b = (er)^{-1} \bmod \lambda(N)$ 
for a random $r$ such that $\gcd(r, \lambda(N)) = 1$. The private key then can be computed as $d = rb \bmod \lambda(N)$. 
In this proposal, it is even suggested that $r$ can be hard coded. We tested our attack 
for a hard coded (known) choice of $r$ and verified that key recovery works in this case. 
Alternatively, if $r$ is not hard coded but we have a trace for the initial $\gcd(r, \lambda(N))$ computation 
using binary gcd, we can again decode it (with the knowledge of $N$) to recover $r$.
With $r$ recovered, the attack proceeds as before, i.e. from the execution trace of $b = (er)^{-1} \bmod \lambda(N)$ 
we recover $p$ and $q$ by running~\cref{alg:branchprunelambda} with $er$ supplied as input instead of $e$. 
Since~\cref{alg:branchprunelambda} is agnostic with respect to the size of $e$, it will handle the 
full size $er$ and recover $p$ and $q$.

\begin{listing}[t]
\begin{lstlisting}[basicstyle=\scriptsize,numbers=left,stepnumber=1]
int wc_ecc_mulmod_ex(mp_int* k, ecc_point *G, ecc_point *R, mp_int* a, mp_int* modulus, int map, void* heap) { ...
for (;;) {
if (--bitcnt == 0) { /* grab next digit as required */
  if (digidx == -1) {
      break;
  }
  buf = get_digit(k, digidx);
  bitcnt = (int)DIGIT_BIT;
  --digidx;
}
i = (buf >> (DIGIT_BIT - 1)) & 1;  /* grab the next msb from the multiplicand */
buf <<= 1;
if (mode == 0) {
  mode = i;  /* timing resistant - dummy operations */
  err = ecc_projective_add_point(M[1], M[2], M[2], a, modulus, mp);...
  err = ecc_projective_dbl_point(M[2], M[3], a, modulus, mp);...
}...
err = ecc_projective_add_point(M[0], M[1], M[i^1], a, modulus, mp);...
err = ecc_projective_dbl_point(M[2], M[2], a, modulus, mp);...
} /* end for */...}
\end{lstlisting}
 \caption{
 \texttt{wc\_ecc\_mulmod\_ex} implements scalar multiplication using a bit-by-bit always-add-and-double algorithm. The function protects against both timing and cache attacks by executing dummy instructions.
 For brevity, error checking and code sections that are not relevant to our attack are removed.
 }
\label{lst:wolfscalmul}
\end{listing}

\subsection{Breaking ECDSA Timing Protection} \label{sec:wolfecdsa}
WolfSSL uses the subroutine \texttt{wc\_ecc\_mulmod\_ex} (\cref{lst:wolfscalmul}) to compute 
the scalar multiplication $k \times G$ while generating the signature. 
This subroutine has built-in mitigations against side-channel attacks and 
implements an \texttt{always-add-and-double} algorithm by arithmetizing the conditional check 
for the \texttt{add}. As a result the scalar operations 
\texttt{add} at Line 15/18 and \text{double} at Line 16/19 will both be executed 
for all scalar bits. This prevents an adversary learning the nonce $k$ bit by bit. 
The second countermeasure that is implemented in this implementation aims to protect 
against attacks exploiting the bit length of the nonce~\cite{moghimi2019tpm,brumley2011remote}. 
This is done by executing a sequence of dummy operations for each leading zero bit. While 
these dummy operations mitigate side channels like data cache attacks, 
page-level attacks, and timing attacks, we can use \attack to distinguish the branch outcome at Line 13 and 
leak the bit length of nonce $k$. 

\para{Recovering Dummy Operations}
We analyze \texttt{wc\_ecc\_mulmod\_ex} using \attack. In this analysis, we count the number 
of instructions that are executed between consecutive accesses to the page that holds 
the \texttt{ecc\_projective\_dbl\_point} subroutine. The trace shows 
that for one transition of basic blocks, we can observe $49$ steps when the function is 
processing the dummy operations. As soon as the subroutine switches to the real operations, 
this step count will change to $46$. As a result, we can use this information to
determine the number of dummy executions of the \texttt{always-add-and-double} sequence from a set of traces.  Since we only need to observe the first few bits in order to recover the length of the nonce, we shortened our trace collection to observe only the first 7 bits.

\para{Lattice Attack using the Nonce Bit Length}
We generated many signature traces, recovered the nonce lengths,  
and filtered for signatures with short nonces~\cite{boneh1996hardness}. 
We followed the approach of Howgrave-Graham and Smart~\cite{howgrave2001lattice} and 
Benger \etal~\cite{benger2014ooh} to formulate the key 
recovery as a lattice problem.

\para{Evaluation}
We executed this attack for 10,000 signing operations. 
Our attack recovered the number of 
leading zero bits with 100\% accuracy. On average, each attack issues 
3244 IRQs to count 2542 steps of the scalar multiplication operation. 
\cref{tbl:cvp_results} shows the results for key recovery using various 
nonce bit lengths. Since the nonce length is recovered without noise, the
lattice attack is quite efficient.

\begin{table}
  \centering
\caption{\label{tbl:cvp_results} Minimum number of signature samples 
for each bias class to reach 100\% recovery success for the lattice-based 
key recovery on \texttt{wc\_ecc\_mulmod\_ex} of ECDSA,
with lattice reduction time {\sc L-Time} and
trace collection time {\sc T-Time}.
}
  \medskip
\resizebox{.87\hsize}{!}{
\begin{tabular}{cccccc}
  \toprule
{\sc LZBs}   & {\sc Dim} &  {\sc L-Time} & {\sc Signatures} &  {\sc IRQs} & {\sc T-Time}\\ \midrule
          4  &        75 &          30 sec &         1,200 &  3.9M  & 13.3 sec \\
          5  &        58 &           5 sec &         1,856 &  6.0M  & 20.4 sec \\
          6  &        46 &           3 sec &         2,944 &  9.6M  & 33.7 sec \\
          7  &        42 &           2 sec &         5,376 &  17.5M & 1  min   \\ \bottomrule
\end{tabular}
}
\end{table}

\section{\attack-Based Side-Channel Analysis} \label{sec:crypto}
Now that we have empirically verified through real-world attacks that \attack 
can recover the runtime control flow of all the branches deterministically, 
we analyze similar cryptographic implementations in other open-source libraries including the latest Libgcrypt 1.8.5, OpenSSL 1.1.1d, and Intel IPP Crypto~\cite{ippcrypto}. 
OpenSSL and Intel IPP Crypto are particularly important for products using 
Intel SGX. Intel has an official wrapper around OpenSSL, called Intel 
SGX-SSL~\cite{sgxssl}. The current version of Intel SGX-SSL is based on 
the stable OpenSSL 1.1.1d. Intel IPP Crypto is the official cryptographic
library by Intel, and it is deployed in many Intel products including Intel SGX SDK~\cite{intelsgxdev}.
\Cref{tab:cryptolist} summarizes our findings in this paper regarding 
vulnerable code paths.

\begin{table*}
  \small
  \centering
  \caption{An overview of applicability of \attack on cryptographic libraries: WolfSSL, Libgcrypt, OpenSSL, IPP Crypto.}
  \medskip
  \label{tab:cryptolist}
  \adjustbox{max width=.99\linewidth}{
  \setlength\tabcolsep{2pt} 
  \begin{tabular}{lllcclc}
    \toprule
    \textbf{}                                     & \textbf{Operation (Subroutine)}                    & \textbf{Implementation}                                        & \begin{tabular}[c]{@{}c@{}}\textbf{Secret} \\ \textbf{Branch}\end{tabular} & \textbf{Exploitable}           & \textbf{Computation} $\rightarrow$ \textbf{Vulnerable Callers}  & \begin{tabular}[c]{@{}c@{}}\textbf{Single-Trace} \\ \textbf{Attack}\end{tabular}           \\
    \midrule
                                                  & Scalar Multiply (\texttt{wc\_ecc\_mulmod\_ex})         & Montgomery Ladder w/ Branches                & \cmark                  & \cmark                     & $(k\times G) \rightarrow$  \texttt{wc\_ecc\_sign\_hash}                                   & \xmark \\\cdashline{2-7}\\
                                                  & Greatest Common Divisor (\texttt{fp\_gcd})                         & Euclidean (Divisions)         & \cmark                  & \xmark                     & N/A                                                                              & N/A                    \\\cdashline{2-7}
                                                  &                                       &                                                                &                         &                            & $(k^{-1} \bmod n) \rightarrow$ \texttt{wc\_DsaSign}                                       & \cmark                    \\
                                                  &                                       &                                                                &                         &                            & $(q^{-1} \bmod p) \rightarrow$ \texttt{wc\_MakeRsaKey}                                    & \cmark                    \\
    \multirow{-5}{*}{\rotatebox{30}{\slshape WolfSSL}}    & \multirow{-3}{*}{Modular Inverse (\texttt{fp\_invmod})} & \multirow{-3}{*}{BEEA}                                         & \multirow{-3}{*}{ \cmark}& \multirow{-3}{*}{ \cmark} & $(e^{-1} \bmod \Lambda(N)) \rightarrow$ \texttt{wc\_MakeRsaKey}                           & \cmark                    \\
    \midrule
                                                    & Greatest Common Divisor (\texttt{mpi\_gcd})                        & Euclidean (Divisions)         & \cmark                  & \xmark                     & N/A                                                                              & N/A                    \\\cdashline{2-7}
                                                    &                                       &                                                                &                         &                            & $(k^{-1} \bmod n) \rightarrow$ \texttt{\{dsa,elgamal\}.c::sign,\_gcry\_ecc\_ecdsa\_sign} & \cmark                    \\
                                                    &                                       &                                                                &                         &                            & $(q^{-1} \bmod p) \rightarrow$ \texttt{generate\_\{std,fips,x931\}}                & \cmark                    \\
                                                  \multirow{-4}{*}{\rotatebox{30}{\slshape Libgcrypt}}  & \multirow{-3}{*}{Modular Inverse (\texttt{mpi\_invm})}  & \multirow{-3}{*}{Modified BEEA~\cite[Vol II, \S4.5.2]{knuth2014art}} & \multirow{-3}{*}{ \cmark}& \multirow{-3}{*}{ \cmark} & $(e^{-1} \bmod \Lambda(N)) \rightarrow$ \texttt{generate\_\{std,fips,x931\}}       & \cmark                    \\
    \midrule
                                                    & Greatest Common Divisor (\texttt{BN\_gcd})                         & BEEA                                                           & \cmark                  & \cmark                     & $gcd(q-1, p-1) \rightarrow$ \texttt{RSA\_X931\_derive\_ex}                                & \cmark                    \\\cdashline{2-7}
                                                  \multirow{-2}{*}{\rotatebox{30}{\slshape OpenSSL}}    & Modular Inverse (\texttt{BN\_mod\_inverse\_no\_branch})& BEEA w/ Branches                                    & \xmark                  & N/A                        & N/A                                                                              & N/A                    \\
    \midrule
                                                    &                                       &                                                                &                         & \qmark                     & $gcd(q-1, e) \rightarrow$ \texttt{cpIsCoPrime}                                            & N/A                  \\
                                                  & \multirow{-2}{*}{Greatest Common Divisor (\texttt{ippsGcd\_BN})}    & \multirow{-2}{*}{Modified Lehmer's GCD}                        & \multirow{-2}{*}{ \cmark}& \qmark                    & $gcd(p-1,q-1) \rightarrow$ \texttt{isValidPriv1\_rsa}                                     & N/A                  \\\cdashline{2-7}
                                              \multirow{-3}{*}{\rotatebox{30}{\slshape IPP Crypto}} & Modular Inverse (\texttt{cpModInv\_BNU})                & Euclidean (Divisions)         & \cmark                  &\xmark                        & N/A                                                                              & N/A                       \\
    \bottomrule
    \end{tabular}
  }
\end{table*}

\subsection{Libgcrypt Analysis}
Libgcrypt uses a custom implementation of the extended Euclidean algorithm  to compute modular inverses (\cref{alg:modinvgcrypt}). 
This algorithm is based on an exercise from The Art of Computer 
Programming~\cite[Vol II, \S4.5.2, Alg X]{knuth2014art}. The algorithm is an 
adaptation of Algorithm X to use the efficient  divide by 2 reduction steps in 
the Binary Euclidean Algorithm. The algorithm computes a vector $(u_1, u_2, u_3)$
such that $uu_1+vu_2 = u_3 = \gcd(u,v)$ using auxiliary vectors $(v_1, v_2, v_3), (t_1, t_2, t_3)$. 
The iterations preserve the invariants $ut_1+vt_2 = t_3$, $uu_1+vu_2=u_3$ 
and $uv_1+vv_2=v_3$. This algorithm is used in numerous
places for secret operations. 

\begin{algorithm}[t]
  \caption{Modular inversion using a variant of BEEA.}
  \label{alg:modinvgcrypt}
  \begin{algorithmic}[1]
    \Procedure{modInv}{$u$, modulus $v$}
    \State $u_1 \gets 1, u_2 \gets 0, u_3 \gets u$
    \State $v_1 \gets v, v_2 \gets u_1 - u, v_3 \gets v$
    \If {$isOdd(u)$} 
      \State $t_1 \gets 0, t_2 \gets -1, t_3 \gets -v$
    \Else
      \State $t_1 \gets 1, t_2 \gets 0, t_3 \gets u$   
    \EndIf
    \While{$t_3 \neq 0$}
      \While{$isEven(t_3)$}
        \If {$isOdd(t_1)$ or $isOdd(t_2)$}
          \State $t_1 \gets t_1 + v, t_2 \gets t_2 - u$
        \EndIf
        \State $t_1 \gets t_1/2,t_2 \gets t_2/2,t_3 \gets t_3/2$
      \EndWhile
      \If {$t_3 > 0$}
        \State $u_1 \gets t_1, u_2 \gets t_2, u_3 \gets t_3$
      \Else
        \State $v_1 \gets v-t_1, v_2 \gets -u-t_2, v_3 \gets -t_3$      
      \EndIf
      \State $t_1 \gets u_1-v_1, t_2 \gets u_2-v_2, t_3 \gets u_3-v_3$
      \IIf {$t_1 < 0$}  $t_1 \gets t_1+v, t_2 \gets t_2-u$  
    \EndWhile
    \Return $u_1$
    \EndProcedure
  \end{algorithmic}
\end{algorithm}

\para{$k^{-1} \bmod n$ in DSA, ECDSA and ElGamal}
The DSA, ECDSA and ElGamal signature schemes all require computing $k^{-1} \bmod n$. 
In Libgcrypt, all of these computations are performed using \cref{alg:modinvgcrypt}.
We derive a single-trace attack similar to \cref{sec:wolfdsa} that recovers all the branches of this algorithm during this computation. 
This trivially leaks $k^{-1}$ for each of these algorithms in a single-trace attack.
As a result, they are all vulnerable to the attack described in \cref{sec:wolfdsa}. 
Note that no masking countermeasure is used for DSA and ElGamal, and we discuss below how the masking countermeasure for ECDSA
is insecure. 

\para{ECDSA Masking Countermeasure}
We identified two vulnerabilities in how masking is applied 
during ECDSA signing in Libgcrypt, as shown in \cref{lst:gcryptecdsamasking}, which leaves it 
vulnerable to attacks against \cref{alg:modinvgcrypt} and a single-trace attack during the computation of $k^{-1} \bmod n$.
Using a randomly chosen blinding variable $b$, Libgcrypt computes the blinded signature
as $s_{b} = k^{-1}(hb + bdr) \bmod n$. To compute the unblinded signature, it computes
$s = s_{b} b^{-1} \bmod n$. The first vulnerability is that
$k^{-1} \bmod n$ is not blinded, so a single-trace attack on this operation simply 
recovers the nonce $k$. This blinding should be modified to $s_b = (k b)^{-1} (h + xr) \bmod n$, 
and this can be unblinded by computing $s = s_b b \bmod n$. 

\begin{listing}[ht]
\begin{lstlisting}[basicstyle=\scriptsize, numbers=left, stepnumber=1]
mpi_mulm (dr, b, skey->d, skey->E.n);
mpi_mulm (dr, dr, r, skey->E.n);     /* dr = d*r mod n (blinded) */
mpi_mulm (sum, b, hash, skey->E.n);
mpi_addm (sum, sum, dr, skey->E.n);  /* sum = hash + (d*r) mod n  (blinded) */
mpi_mulm (sum, bi, sum, skey->E.n);  /* undo blinding by b^-1 */
mpi_invm (k_1, k, skey->E.n);        /* k_1 = k^(-1) mod n  */
mpi_mulm (s, k_1, sum, skey->E.n);   /* s = k^(-1)*(hash+(d*r)) mod n */
\end{lstlisting}
  \vspace{-.5em}
  \caption{
    The masking protection for ECDSA leaves the $k^{-1} \bmod n$ operation vulnerable to our 
    single-trace attack.
  }
\label{lst:gcryptecdsamasking}
\end{listing}

The second vulnerability is that since $b$ needs to be inverted in this blinding scheme, 
Libgcrypt computes the $b^{-1} \bmod n$ using the same vulnerable implementation (\cref{lst:gcryptecdsablind}). 
Therefore, a single-trace attack can also recover the blinding value. 

\begin{listing}[ht]
\begin{lstlisting}[basicstyle=\scriptsize, numbers=left, stepnumber=1]
do {    _gcry_mpi_randomize (b, qbits, GCRY_WEAK_RANDOM);
        mpi_mod (b, b, skey->E.n);
} while (!mpi_invm (bi, b, skey->E.n));
\end{lstlisting}
  \vspace{-.5em}
  \caption{
    \texttt{\_gcry\_ecc\_ecdsa\_sign} computes the modular inverse of the blinding factor $b$ using a vulnerable function.
  }
\label{lst:gcryptecdsablind}
\end{listing}

\para{RSA Input Masking}
To avoid timing attacks, RSA decryption and signing in Libgcrypt use masking on the 
input ciphertext or message. For a random variable $r$ and input ciphertext $c$, the decryption is performed on 
$m_b = (c r^e)^d \bmod n = c^d r \bmod n$. The message can then be unblinded by computing $m = m_b r^{-1} = c^d \bmod n$.
Unfortunately, the $r^{-1} \bmod n$ is also computed using the vulnerable modular inverse function. 
As a result, a single-trace attack can recover the blinding factor, rendering this countermeasure ineffective. 

\para{RSA Key Generation}
Three RSA key generation subroutines in Libgcrypt: \texttt{generate\_std}, \texttt{generate\_fips} and \texttt{generate\_x931} all
use the vulnerable \texttt{mpi\_invm} function to compute both $q^{-1} \bmod p$ and $e^{-1} \bmod \lambda(N)$, and are vulnerable to the attacks described in \cref{sec:wolfrsa}.

\begin{listing}[ht]
\begin{lstlisting}[basicstyle=\scriptsize, numbers=left, stepnumber=1]
if (!BN_sub(r1, rsa->p, BN_value_one())) goto err;     /* p-1 */
if (!BN_sub(r2, rsa->q, BN_value_one())) goto err;     /* q-1 */
if (!BN_mul(r0, r1, r2, ctx)) goto err;                /* (p-1)(q-1) */
if (!BN_gcd(r3, r1, r2, ctx)) goto err;
\end{lstlisting}
  \vspace{-.5em}
  \caption{
    \texttt{RSA\_X931\_derive\_ex} uses \texttt{BN\_gcd} to compute $\lambda(N)$, exposing $p$ and $q$
    to our attack.
  }
\label{lst:opensslx931}
\end{listing}

\subsection{Analysis of OpenSSL}
After many iterations and multiple attacks~\cite{garcia2017constant,weiser2018single}, 
OpenSSL implemented a constant-time modular inversion function, \texttt{BN\_mod\_inverse\_no\_branch} 
for DSA, ECDSA, and RSA key generation. 
In various critical primitives, this function is also used to compute the GCD. 
However the legacy binary GCD function is still supported in the latest OpenSSL 
code base, version 1.1.1d, in the function \texttt{BN\_gcd} (\cf Appendix \cref{alg:gcdopenssl}). 
The subroutine \texttt{RSA\_X931\_derive\_ex}, 
which is responsible for generating RSA keys according to the X.931 standard, 
uses this function during the computation of $\lambda(N)= \lcm(p-1,q-1) = (p-1)(q-1)/gcd(p-1, q-1)$, as shown in \cref{lst:opensslx931}. 
Thus we can apply our attack technique from \cref{sec:wolfrsa} to recover the RSA private key 
from the computation of $gcd(p-1, q-1)$.

\subsection{Analysis of Intel IPP Crypto}
The \emph{Intel IPP Crypto} library uses a conventional Euclidean algorithm to compute modular inverses.
This algorithm performs a series of division operations in a loop. While \attack can recover 
the precise number of division operations, this leakage does not seem to be exploitable during the RSA key generation~\cite[\S6]{moghimi2019tpm}.

On the other hand, for computing the GCD, Intel IPP Crypto uses a modified version of Lehmer's GCD algorithm~\cite{sorenson1995analysis}. Lehmer's GCD algorithm 
and Intel's modified implementation are not constant time, and have secret-dependent branches~\cite{intellehmergcd}.
This GCD implementation is only used during RSA key generation, where only a single-trace attack results in a vulnerability.
Our analysis in \cref{sec:wolfrsa} does not directly apply to this algorithm, and we leave the analysis and potential exploitability of this implementation for future work. 
This potential oversight in Intel's GCD implementation once more illustrates the intricacies of applying Intel's own recommended constant-time programming guidelines~\cite{IntelCst}.

\subsection{More Single-Trace Attack Evaluations}
\para{Attack on DSA, ECDSA and ElGamal (Libgcrypt)}
We replicated the attack in \cref{sec:wolfdsa} using synthetic traces from~\cref{alg:modinvgcrypt}.
We ran the attack on 100 different $k^{-1} \bmod n$ and recovered $k_{inv}$ and the secret key in all cases.  The attack applies to ElGamal as well by computing the private key $x = r^{-1}(h-s k) \bmod (p-1)$. 

\para{Attack on RSA Key Generation (Libgcrypt, OpenSSL)}
We replicated synthetic traces of branches from OpenSSL's binary GCD algorithm executed on $\gcd(q-1, p-1)$. We applied \cref{alg:branchprune} with a modified test function modeling this algorithm, and applied a heuristic to match the appropriate number of trace steps to the bits guessed so far. 
We ran the attack using synthetic traces for $100$ different 256-bit RSA keys. This key size is chosen to efficiently verify the correctness of our algorithm. Our attack successfully recovered every key.
We similarly replicated the same attack as \Cref{sec:wolfrsa} with a test function
following \cref{alg:modinvgcrypt}. Similarly, we ran the attack using synthetic traces for $100$ different 256-bit RSA keys and the attack was successful in all cases.

\section{Limitations and Future Work}
\label{sec:limit}

\para{Vulnerable Code Patterns}
\attack interrupts a victim enclave precisely one instruction at a time and
relies on a secondary page-table oracle to assign a spatial resolution to each
instruction-granular observation.
Our attack is thus only effective when the victim code contains a
secret-dependent branch that accesses a different
code or data page at the same instruction offset in both execution paths.
In contrast to previous controlled-channel
attacks~\cite{xu2015controlled,shinde2016preventing,van2017telling}, our notion
of instruction-granular page access traces allows the \emph{sequence} of code
and data page visits in both branches to be identical.

We essentially only need a \enquote{marker} page that is accessed at a different
relative instruction offset in the secret-dependent execution path.
We found that in practice compilers generate code with different page accesses
at different instruction offsets in both branches for a variety of reasons,
including data or stack accesses, arithmetic operations, and subroutine calls.
The evaluation of the previous sections clearly shows that the
instruction-granular page access traces extracted by \attack are strictly
stronger, and can hence target more vulnerable code patterns, than the page
fault sequences exploited by prior controlled-channel attacks.
In order to not be vulnerable to \attack, secret-dependent code paths
should ideally be avoided altogether, or they should be explicitly aligned in
such a way that both branches always access the exact same set of code and data
pages for every instruction among both execution paths.

\para{Automation Opportunities}
The case-study attacks presented in this paper relied on careful manual
inspection of the victim enclave source code and binary layout to identify
vulnerable secret-dependent code patterns.
We expect that dynamic analysis and symbolic execution approaches could further
improve the effectiveness of our attacks, and increase confidence for defenders,
by automating the discovery of vulnerable code patterns~\cite{wichelmann2018microwalk,wang2017cached} 
and possibly even the synthesis of proof-of-concept exploitation code.
While the requirements for vulnerable code patterns are relatively clear-cut, as
described above, we expect that it may be particularly challenging to
automatically track the propagation of secrets and distinguish between
non-secret and secret-dependent control flows~\cite{blazy2019verifying}.

\para{Comparison to Branch Prediction Leakage}
\Cref{tab:sgx-attack} identified branch prediction side
channels~\cite{lee2017inferring,evtyushkin2018branchscope,huo2020bluethunder}
as an alternative attack vector to spy on enclave control flow at an
instruction-level granularity with reasonable accuracy.
In contrast to \attack, however, microarchitectural leakage from branch
predictors is inherently noisy and typically requires multiple runs of the
victim enclave, ruling out this class of side channels for the noiseless
single-trace attacks on key generation algorithms presented in this paper.
Furthermore, in contrast to the architectural interrupt and paging interfaces
exploited by \attack, branch prediction side-channel leakage can be eradicated
relatively straightforwardly by flushing branch history buffers when exiting the
enclave, similar to the microcode updates Intel already distributed to
flush branch predictors on enclave entry in response to Spectre
threats~\cite{chen2018sgxpectre}.

In addition to being deterministic, our attack is significantly easier to scale
and replicate, considering that branch predictors feature a complex design that
may change from one microarchitecture to the other.
BranchScope~\cite{evtyushkin2018branchscope}, for instance, relies on finding a
heuristic through reverse engineering to probe a specific branch.
This heuristic is dependent on \begin{paraenum}
    \item the state of other components like global and tournament predictors; and 
    \item the exact binary layout of the victim program.
\end{paraenum}
Previous attacks focus on distinguishing one or a small number of branches and we believe
that replicating BranchScope to probe multiple branches across various targets
(\eg BEEA) would be challenging and may even be practically infeasible.
\attack, in contrast, is much easier to replicate, and we showed in \cref{sec:wolfssl}
that our attack scales to probing the entire execution path in a single run.

\section{Mitigation Strategies} \label{sec:countermeasure}

\para{Interrupt Detection}
\attack relies on the ability to single-step enclaved execution, which is within Intel SGX's threat model~\cite{inteSDM,vanbulck2017SGXStep}.
While SGX enclaves remain explicitly interrupt-unaware by design, some research proposals~\cite{shih2017t,chen2017detecting} retrofit hardware support for transactional memory to detect suspicious interrupt rates as a side-effect of an ongoing attack.
However, such features are not commonly available on off-the-shelf SGX hardware, and they would not fundamentally address the attack surface as \attack adversaries are likely to develop stealthier attack techniques~\cite{van2017telling,wang2017leaky} that remain under the radar of heuristic defenses.
Nevertheless, following a long line of microarchitectural attacks~\cite{moghimi2017cachezoom,van2018nemesis,vanbulck2017SGXStep,huo2020bluethunder,lee2017inferring} abusing interrupts, our study provides strong evidence that interrupts may also amplify deterministic controlled-channel leakage and should be taken into account in the enclaved execution threat model.
We advocate for architectural changes in the Intel SGX design and further research to rule out interrupt-driven attack surface~\cite{busi2020provably}.

\para{Self-Paging}
Recent work~\cite{orenbach2020autarky} proposes modifications to the
Intel SGX architecture to rule out page-fault controlled channels by
delegating paging decisions to the enclave.
The proposed design modifies the processor to no longer report the faulting page
base address to the untrusted OS and to not update \enquote{accessed} and
\enquote{dirty} page-table attributes when in enclave mode.
While these modifications would indeed thwart the deterministic spatial
dimension of the \attack instantiations described in this paper, 
we expect that adversaries may adapt by resorting to alternative side-channel
oracles to construct instruction-granular page access patterns.
A particularly promising avenue in this respect would be to combine \attack
interrupt counting with the distinct timing differences observed for unprotected
page-table entries that were brought into the CPU cache during enclaved
execution~\cite{van2017telling}.

\para{Static Code Balancing}
We encourage future research in improved compile-time hardening techniques that may automatically rewrite conditional branches to always ensure a constant interrupt counting pattern, regardless of the executed code.
The key requirement for such a defense would be to ensure that the adversary not only observes a secret-independent sequence of pages but also always counts the same number of instructions between page transitions.
To achieve such a guarantee, the compiler would also have to be explicitly aware of 
macro fusion, as explained in \cref{sec:fusion}, when balancing the observed instruction counts.
We expect further challenges when dealing with secret-dependent loop bounds, 
as in our attacks of \cref{sec:wolfssl}.
To handle data accesses, control-flow balancing techniques could potentially be combined with existing solutions for data location randomization~\cite{brasser2019dr}.

\para{Constant-Time Implementation}
The best practice for cryptographic implementations is to avoid secret-dependent branches and memory lookups. 
WolfSSL applied such a countermeasure to mitigate our attack on ECDSA (CVE-2019-19960). 
Bernstein and Yang proposed a constant-time GCD algorithm that can be used for applications like modular inversion~\cite{bernstein2019fast}. 
However, constant-time implementations are not easy for generic non-cryptographic applications~\cite{stefanov2013path}. 
While there are tools and techniques to test~\cite{wichelmann2018microwalk}, verify~\cite{almeida2016verifying}, and generate~\cite{bond2017vale} constant-time code, the scalability and performance of these approaches is still far from settled.

\para{Cryptographic Countermeasures}
While it is preferable to avoid secret-dependent branches altogether, 
specific countermeasures can be applied to some cryptographic schemes. 
As we discuss in \cref{sec:crypto}, masking the input of the modular inversion
can mitigate our demonstrated attack if it is applied properly and the blinding value itself is not leaked. 
WolfSSL applies this solution to mitigate our attack on DSA (CVE-2019-19963). 
However, as we showed in \cref{sec:crypto} and \cref{sec:wolfrsa}, these countermeasures should be applied carefully in the presence of a powerful adversary. 

Some operations have more secure alternative implementations. In particular, for the attack on $q^{-1} \bmod p$ 
RSA-CRT key generation (CVE-2020-7960), 
Fermat's Little Theorem computes $q^{p-2} \bmod p$. As a result, the implementation can avoid modular inversion for this operation, and instead rely on a constant-time modular exponentiation implementation.

\section{Conclusion}

Our works show that deterministic controlled-channel adversaries are not restricted 
to observing enclave memory accesses at the level of coarse-grained 4\,KiB pages, 
but can also precisely reconstruct intra-page control flow at a maximal, instruction-level granularity.
We demonstrated the practicality and improved resolution of \attack by discovering highly dangerous 
single-trace key extraction attacks in several real-world, side-channel hardened cryptographic libraries.
In contrast to known microarchitectural leakage sources, the more fundamental threat of 
deterministic controlled-channel leakage cannot be dealt with by merely flushing or partitioning 
microarchitectural state and instead requires research into more principled solutions.

\ifAnon
\else
\section*{Acknowledgments}
We thank our reviewers
for their suggestions that helped improving the paper.
This work is partially funded by the Research Fund KU Leuven, a generous gift from Intel, and by the US National 
Science Foundation under grants no. 1513671, 1651344, 1814406, and 1913210. 
Jo Van Bulck is supported by a grant of the Research Foundation -- Flanders (FWO). 
\fi

\ifUsenix
  \setlength{\bibsep}{1.0pt}
  \bibliographystyle{plain}
  {\footnotesize
  \bibliography{main}
  }
\else
  \bibliographystyle{IEEEtranS}
  {\footnotesize
  \bibliography{IEEEabrv,main}
  }
\fi

\appendix
\section{Appendix}
\subsection{OpenSSL GCD Algorithm} \label{sec:gcdopenssl}
\cref{alg:gcdopenssl} shows the binary GCD algorithm in OpenSSL. 

\begin{algorithm}[ht]
  \caption{OpenSSL Binary GCD Algorithm.}
  \label{alg:gcdopenssl}
  \begin{algorithmic}[1]
    \Procedure{GCD}{$a$, $b$}
    \State $s \gets 0$
    \IIf{$a<b$} $a, b \gets b, a$
    \While{$b \neq 0$}
      \If{$isOdd(a)$}
        \If{$isOdd(b)$}                     \Comment{a is odd, b is odd}
            \State $a \gets a-b, a \gets a/2$         
            \IIf{$a<b$} $a, b \gets b, a$
        \Else                               \Comment{a is odd, b is even}
            \State $b \gets b/2$
            \IIf{$a<b$} $a, b \gets b, a$
        \EndIf              
      \Else
        \If{$isOdd(b)$}                     \Comment{a is even, b is odd}
          \State $a \gets a/2$
          \IIf{$a<b$} $a, b \gets b, a$
        \Else                               \Comment{a is even, b is even}
          \State $a \gets a/2, b \gets b/2, s \gets s+1$              
        \EndIf
      \EndIf
    \EndWhile
    \IIf{$s > 0$} $a \gets a * (2^{s})$
    \State \Return $a$
    \EndProcedure
  \end{algorithmic}  
\end{algorithm}  

\subsection{Branch Shadow-Resistant Code Attack}
\label{app:zigzag}

\Cref{lst:zigzagc} provides an elementary example function with secret-dependent branches.
We provide the corresponding assembly output in \cref{lst:zigzag}, 
as produced by the LLVM-based, open-source compiler mitigation pass~\cite{hosseinzadeh2018mitigating} 
against branch shadowing attacks, described in \cref{sec:zigzag}.
We passed the \texttt{-mllvm -x86-branch-conversion} and \texttt{-mllvm -x86-bc-dummy-instr} options 
to enable both rewriting of conditional branches via the trampoline area and protection against 
timing attacks via dummy instruction balancing.
Note that the randomizer is not integrated in the open-source release, and all code blocks on 
the trampoline area would still have to be randomly re-shuffled at runtime to protect against branch-shadowing attacks. To achieve sufficient entropy, trampoline areas have to be larger than 4\,KiB~\cite{hosseinzadeh2018mitigating}, and  hence the trampoline will occupy at least one separate page.

We reveal control flow in the instrumented code of \cref{lst:zigzag} using \attack as follows.
In the case where the secret-dependent if condition is true, the indirect branch at line~\ref{lbl:cmove} will execute 
the single-instruction \texttt{jmp\_if} block on the trampoline page, followed by 4 instructions on 
the instrumented code page, totaling 5 instructions before reaching the \texttt{end\_if} marker.
In contrast, if the if condition is false, the indirect branch at line~\ref{lbl:cmove} will 
transfer to the \texttt{skip\_if} block on the trampoline page, totaling 4 instructions before 
eventually reaching the \texttt{end\_if} marker back on the instrumented code page.
Similar unbalanced instruction counts follow for the else block.

We experimentally verified that \attack adversaries can deterministically learn the if condition 
by merely counting instructions and observing page accesses.
Moreover, because the dummy instructions do not result in exactly balanced instruction counts, as explained above, merely counting the total amount of executed instructions even suffices in itself without having to distinguish accesses to the trampoline page.

\begin{listing}[t!]
\begin{lstlisting}
void my_func(int a) {
    if (a  != 0) block1++; else block2++;
    block3++;
}
\end{lstlisting}
  \caption{
    Sample code snippet with conditional branching.
  }
\label{lst:zigzagc}
\end{listing}

\begin{listing}[t!]
\begin{lstlisting}[style=customasm,numbers=left, stepnumber=1]
          jmp    my_func /***  BEGIN TRAMPOLINE ***/
jmp_done: jmp    done
jmp_done2:jmp    done
skip_else:add    $0x0,%r13b   # compensating dummy
          lea    jmp_done2(%rip),%r15
          jmp    end_else
jmp_else: jmp    else
skip_if:  add    $0x0,%r13b   # compensating dummy
          add    $0x0,%r13b   # compensating dummy
          lea    jmp_else(%rip),%r15
          jmp    end_if
jmp_if:   jmp    if      /***  END TRAMPOLINE ***/
my_func:  push   %rbp
          mov    %rsp,%rbp
          mov    %edi,-0x4(%rbp)
          cmpl   $0x0,-0x4(%rbp)
          lea    jmp_if(%rip),%r15
          lea    skip_if(%rip),%r13
          cmove  %r13,%r15
        @\label{lbl:cmove}@  jmp    *%r15
if:       mov    block1(%rip),%eax
          add    $0x1,%eax
          mov    %eax,block1(%rip)
          lea    skip_else(%rip),%r15
end_if:   jmp    *%r15
else:     mov    block2(%rip),%eax
          add    $0x1,%eax
          mov    %eax,block2(%rip)
          lea    jmp_done(%rip),%r15
end_else: jmp    *%r15
done:     mov    block3(%rip),%eax
          add    $0x1,%eax
          mov    %eax,block3(%rip)
          pop    %rbp
          ret
\end{lstlisting}
  \caption{
    Hardened assembly output, corresponding to the source code in \cref{lst:zigzagc}, as produced by the open-source branch shadowing mitigation LLVM compiler pass.
  }
\label{lst:zigzag}
\end{listing}

\end{document}